\documentclass[aps,pra,twocolumn,eqsecnum,amsmath,superscriptaddress,showpacs]{revtex4}

\usepackage{graphicx}
\usepackage{longtable}
\usepackage{epsfig}
\usepackage{dcolumn}
\usepackage{bm}
\usepackage{amssymb}
\usepackage{multirow}
\usepackage{times,color}
\begin{document}

\title{Quantum phase transitions in exactly solvable one-dimensional compass models}

\author {    Wen-Long You }
\affiliation{Max-Planck-Institut f\"ur Festk\"orperforschung,
             Heisenbergstrasse 1, D-70569 Stuttgart, Germany }
\affiliation{School of Physical Science and Technology, Soochow University,
             Suzhou, Jiangsu 215006, People's Republic of China  }

\author {    Peter Horsch }
\affiliation{Max-Planck-Institut f\"ur Festk\"orperforschung,
             Heisenbergstrasse 1, D-70569 Stuttgart, Germany }

\author {    Andrzej M. Ole\'s }
\affiliation{Max-Planck-Institut f\"ur Festk\"orperforschung,
             Heisenbergstrasse 1, D-70569 Stuttgart, Germany }
\affiliation{Marian Smoluchowski Institute of Physics, Jagellonian
             University, Reymonta 4, PL-30059 Krak\'ow, Poland }

\date{4 December 2013}

\begin{abstract}
We present an exact solution for a class of one-dimensional compass
models which stand for interacting orbital degrees of freedom in a Mott
insulator. By employing the Jordan-Wigner transformation we map these
models on noninteracting fermions and discuss how spin correlations,
high degeneracy of the ground state, and $Z_2$ symmetry in the quantum
compass model are visible in the fermionic language. Considering a
zigzag chain of ions with singly occupied $e_g$ orbitals ($e_g$ orbital
model) we demonstrate that the orbital excitations change qualitatively
with increasing transverse field, and that the excitation gap closes at
the quantum phase transition to a polarized state. This phase
transition disappears in the quantum compass model with maximally
frustrated orbital interactions which resembles the Kitaev model.
Here we find that finite transverse field
destabilizes the orbital-liquid ground state with macroscopic
degeneracy, and leads to peculiar behavior of the specific heat and
orbital susceptibility at finite temperature. We show that the entropy
and the cooling rate at finite temperature exhibit quite different
behavior near the critical point for these two models.
\end{abstract}

\pacs{75.10.Jm, 05.30.Rt, 75.25.Dk, 75.40.Cx}
\maketitle

\section{Introduction}
\label{sec:intro}

In recent years the growing interest in orbital degrees of freedom
for strongly correlated electrons in transition-metal oxides
(TMOs) \cite{Kug82,Fei97,Tokura,Hfm}, was amplified by complex
phenomena uncovered in theory and experiment, such as the interplay
between spin and orbital degrees of freedom \cite{Ole05,Kha05,Ole12},
consequences of orbital degeneracy in the perovskite vanadates
\cite{Hor03}, phase transitions to magnetic and orbital order
\cite{Hor08}, dimerization in ferromagnetic spin-orbital chains
\cite{Her11}, entanglement entropy spectra in one-dimensional (1D)
models,\cite{You12}, and exotic types of spin order triggered by
spin-orbital entanglement in the Kugel-Khomskii models \cite{Brz13}.
Electrons are strongly correlated and localize due to large on-site
Coulomb interaction $U$ --- then they interact by superexchange.
While spin and orbital degrees of freedom are generally entangled and
influence each other on superexchange bonds \cite{Ole12,You12,Brz14},
or due to local spin-orbit coupling \cite{Jac09,Cha10}.
In spin-orbital systems an electron can break into a spinon and an
orbiton \cite{Woh11}, as observed recently in Sr$_2$CuO$_3$ \cite{Sch12}.
This motivates a more careful study of orbital models in low dimension.
Such models for Mott insulators, depend on the type of partly filled
$3d$ orbitals, with either $e_g$ symmetry \cite{vdB99,vdB04,Fei05,Ryn10},
or $t_{2g}$ symmetry \cite{Dag08,Wro10,Tro13,Che13}.

In TMOs with the perovskite structure active orbitals are selected by
the octahedral crystal field due to the oxygen ions which splits the
$3d$ quintet at a transition-metal ion into a $t_{2g}$ triplet and an
$e_g$ doublet at higher energy. Well known examples of $e_g$ systems
with partly filled $e_g$ orbitals by one spin flavor which are of
interest here are:
(i) $d^4$ ions (in LaMnO$_3$, Rb$_2$CrCl$_4$, or KCrF$_3$) \cite{Fei99},
(ii) $d^7$ ions in LiNiO$_2$ \cite{Rei05}, or
(iii) $d^9$ ions \cite{Kug82} in KCuF$_3$, K$_3$Cu$_2$F$_7$, or
K$_2$CuF$_4$ \cite{Brz13}.
In all these systems the $t_{2g}$ orbitals are either completely filled
(in the $d^7$ and $d^9$ configurations), or contain one electron each
(in the $d^4$ configuration) -- in the latter case their spins are
aligned with the spin of an $e_g$ electron due to Hund's exchange.
The two $e_g$ orbitals represent then the dynamical degrees of freedom.

Here we focus on ferromagnetic states with spins fully polarized
were only the orbital degrees of freedom being active. Orbitals are
interacting via generically anisotropic superexchange interactions
$\propto J_{\gamma}$ depending on the bond direction $\gamma=a,b$.
Thus a typical orbital superexchange model has the following
anisotropic form,
\begin{equation}
H_J= \sum_{\langle ij\rangle\parallel\gamma}
J_{\gamma}^{} T_{i}^{\gamma}T_{j}^{\gamma}.
\label{Horb}
\end{equation}
This model stands for intrinsically frustrated {\it directional\/}
orbital interactions on the square lattice, and may represent both
$e_g$ \cite{vdB99} and $t_{2g}$ orbital interactions \cite{Dag08}.
In the latter case the operators include just one of the orthogonal
pseudospin components at each bond and are Ising-like. This form of
interactions is found as well in the compass models
\cite{vdB13,Nus05,Dou05,Dor05,Chen07,Wen08,Orus09,Cin10,Brz10,Tro10,Brz07,Brz09},
and in the Kitaev model on the honeycomb lattice
\cite{Kit06,Che08,Div09}.

The interactions that are considered here are defined by the pseudospin
operators $T^{\gamma}_i$ for two active orbitals (for $T=1/2$), and we
define them as linear combinations of the Pauli matrices
$\{\sigma_{i}^x,\sigma_{i}^y\}$ representing the two pseudospin
components on odd/even bonds \cite{notesz},
\begin{equation}
\tilde{\sigma}_{i}(\pm\theta)\equiv
\cos(\pm\theta/2)\,\sigma_{i}^x +\sin(\pm\theta/2)\,\sigma_{i}^y.
\label{orb}
\end{equation}
These operators define the generalized compass model (GCM) considered
in this paper. In the 1D GCM the interactions depend on the $x$-th and
$y$-th orbital component in Eq. (\ref{orb}), but the exchange
interactions are bond dependent as in Eq. (\ref{Horb}) and alternate
between even ($J_{\rm e}$) and odd ($J_{\rm o}$) exchange bonds along
the 1D chain of $N$ sites (we consider periodic boundary conditions,
and even values of $N$),
\begin{equation}
H_J(\theta)= \sum_{i=1}^{N/2}\left\{
 J_{\rm o}\tilde{\sigma}_{2i-1}( \theta)\tilde{\sigma}_{2i  }( \theta)
+J_{\rm e}\tilde{\sigma}_{2i  }(-\theta)\tilde{\sigma}_{2i+1}(-\theta)\right\},
\label{gJ}
\end{equation}
where we sum over unit cells.
For a representative pseudospin $\tilde{\sigma}_{i}$ the interaction
involves the quantization axis with direction $\theta$ for one bond and
the one with $-\theta$ for the other, so each pseudospin has to find
some compromise. This frustration increases gradually with increasing
angle $\theta$ when the model Eq. (\ref{gJ}) interpolates between the
Ising model at $\theta=0$ to the quantum compass model (QCM) at
$\theta=\pi/2$ \cite{Cin10}. The latter is also called the 1D Kitaev
model by some authors \cite{Div09}. In the intermediate case,
$\theta=\pi/3$, one finds orbital superexchange (\ref{Horb}) for the
$e_g$ orbital model (EOM) or $60^\circ$ compass model
(for the angle $\theta=\pi/3$).

The EOM (at $\theta=60^\circ$) was first introduced as an effective
model for perovskite $e_g$ orbital systems \cite{vdB99}, and next
considered in two-dimensional (2D) and three-dimensional (3D)
ferromagnetic TMOs with active $e_g$ orbitals \cite{Ole05,vdB99,Nus04,Bis05}.
The equivalent planar model describes the insulating phase of $p$-band
fermions in triangular, honeycomb and kagome optical lattices
\cite{Zhao08,Wu08}.

The QCM arises from the GCM Eq. (\ref{gJ}) with frustrated Ising-like
interactions tuned by an angle $\theta$ on a square lattice \cite{Cin10}
at $\theta=90^\circ$. While 2D Ising models with frustrated
interactions have long-range order at finite temperature \cite{Lon80},
one might expect that disordered states emerge when interacting spin
components depend on the bond direction, as in Eq. (\ref{Horb}).
This is indeed the case of the Kitaev model on a hexagonal lattice
with a spin-liquid ground state that is exactly solvable \cite{Kit06}.
Instead, the infinite degeneracy in the ground state for the classical
compass model on 2D or 3D cubic lattices is lifted via the
order-out-of-disorder mechanism and a directional ordering of
fluctuations appears at low temperature \cite{Mis04}. For the quantum
version, it has been rigorously proven in terms of the reflection
positivity method \cite{You07} that the alternating orbital order is
stable in the 2D planar $60^\circ$ compass model at zero temperature.
Indeed, this result is confirmed by numerical simulations \cite{Wen08}.

The QCM is characterized by an exotic property of the dimensional
reduction which implies that a $d$-dimensional system has long-range
order in $(d-1)$ dimension \cite{vdB13,You08}. For example, the global
ground states of the 2D QCM have a ferro-orbital nematic long-range
order in a highly degenerate ground state \cite{Brz10,You10,He11}.
It has been shown that this directional long-range order survives in a
manifold of low energy excited states when the compass interactions are
perturbed by the Heisenberg ones \cite{Tro10} ---
this property opens its potential application in quantum computation.
It is remarkable that the 2D QCM is dual to the toric code model in
transverse magnetic field \cite{Vid09} and to the Xu-Moore model
(Josephson arrays) \cite{Xu04}.

Quantum phase transitions (QPTs) between different types of order
were established in the 1D QCM \cite{Brz07}, in a quantum compass
ladder \cite{Brz09}, and in the 2D QCM
\cite{Nus05,Dou05,Dor05,Chen07,Wen08,Orus09,Cin10,Brz10,Tro10},
when anisotropic interactions are varied through the isotropic point
and the ground state switches between two different types of Ising
nematic order dictated by either interaction. At the transition point
itself, i.e., when the competing interactions are balanced, the ground
state is highly degenerate and contains states which correspond to both
relevant kinds of nematic order. The correlations along perpendicular
direction to that of the nematic order are restricted to nearest
neighbor (NN) sites \cite{You11}, and certain NN spin correlations
change discontinuously at the critical point. Studies of the 1D QCM
using entanglement measures and quantum discord in the ground state show
that the correlations between two orbitals on some bonds are essentially
classical \cite{You12b}. The QPT driven by the transverse field emerges
only at zero field and is of the second order \cite{Sun09}.

The purpose of this paper is to present an exact solution of the GCMs
(with orbitals of $e_g$ or $t_{2g}$ symmetry), and to investigate their
properties at finite temperature. We propose a possible scenario
provided by a 1D zigzag lattice which can be prepared in layered
structures of TMOs \cite{Xiao11}, or are realized in optical lattices
by fermions occupying $p_x$ and $p_y$ orbitals \cite{Simon11,Gsun}.
Our motivation is twofold:
On one hand, recently artificial heterostructures of TMOs are becoming
available, and the modern technologies and allow to devise artificial
1D quantum systems, such as quantum wires or rings. In terms of
interface engineering, some models can be designed, such as a 2D
design for man-made honeycomb lattice \cite{Xiao11}. On the other hand,
the zigzag chain of $S=1/2$ spins, with active
$xy$ and $yz$ orbitals in $d^1$ states at Ti$^{3+}$ ions, is found in
pyroxene titanium oxides ATiSi$_2$O$_6$ (A = Na,Li) \cite{Kon04,Hik04}.
The alternation of the Ti-Ti distance is a direct consequence of
orbital dimerization. We also solve exactly the GCM at arbitrary angle
$\theta$ and compare its properties with those of the EOM. We find that
the EOM and the 1D QCM are both characterized by a QPT, but we uncover
an important difference between these transitions which is found for
the anisotropic interactions.

The paper is organized as follows: We introduce the EOM in Sec.
\ref{sec:eg} and present its exact solution in Sec. \ref{sec:dia}
obtained using the Jordan-Wigner transformation.
We show that a gap found in the excitation spectrum persists also in
the entire range of angle $\theta$ in the GCM, see Sec. \ref{sec:model}.
Properties of the GCM, including the dependence of transverse orbital
polarization and intersite pseudospin correlations on the angle $\theta$
and on the polarizing field are investigated in Sec. \ref{sec:corr}.
This field is responsible for the switch of the pseudospin order at the
QPT. Next we present exact results at finite temperature obtained for
the entropy and for the orbital cooling rate in Sec. \ref{sec:s}, and
for the specific heat in Sec. \ref{sec:cv}. The orbital polarization
induced by finite field and orbital susceptibility are analyzed in Sec.
\ref{sec:chi}. The paper is concluded with a final discussion and
summary in Sec. \ref{sec:summa}.
Here we also highlight the interpretation of the results in terms of
fermionic bands as equivalent to the spin correlations. These
correlations follow from the $Z_2$ symmetry, as explained in
the Appendix.

\section{Orbital compass model}
\label{sec:como}

\subsection{One-dimensional zigzag $e_g$ orbital model}
\label{sec:eg}

\begin{figure}[t!]
\includegraphics[width=8cm]{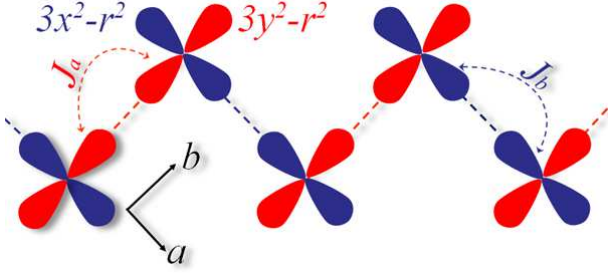}
\caption{(Color online)
Zigzag chain in an $(a,b)$ plane with one hole (or electron) per site
in $e_g$ states of a Mott insulator. The directional orbitals,
$3x^2-r^2$ and $3y^2-r^2$, are the most convenient choice to calculate
the orbital superexchange interactions along the $a$ and $b$ cubic axis,
respectively. In the usual $e_g$ orthogonal orbital basis
$\{3z^2-r^2,x^2-y^2\}$ such interactions may cause orbital flips as the
orbital flavor is not conserved in intersite hopping processes (charge
excitations) \cite{vdB99}. If two $t_{2g}$ orbitals, $zx$ and $yz$,
are considered instead, only diagonal hopping between pairs of these
orbitals occurs along the bonds parallel to the $a$ or $b$ axis
\cite{Dag08}, and one finds the 1D quantum compass model \cite{Brz07}. }
\label{fig:chain}
\end{figure}

We consider first the exact solution for the 1D EOM (60$^\circ$ compass
model) of Fig. \ref{fig:chain}, with the Hamiltonian $H_J$ given by Eq.
(\ref{gcom}) at $\theta=\pi/3$. This example serves as a general
guideline for the analytic solution and for the thermodynamics presented
below in Secs. \ref{sec:gen}-\ref{sec:qpt}. The interactions in Eq.
(\ref{Horb}) are given by operators
\begin{equation}
T_{i}^{a(b)}=
-\frac{1}{2}\sigma_{i}^y \pm \frac{\sqrt{3}}{2} \sigma_{i}^x,
\label{T}
\end{equation}
which depend on Pauli matrices, $\{\sigma_i^{\alpha}\}$ ($\alpha=x,y$)
for $e_g$ orbital states \cite{notesz}. In the case of a 3D cubic system
Eq. (\ref{T}) would be augmented by $T_i^c=\sigma_i^y$ for the bonds
along the $c$ axis. The interactions follow from the Kugel-Khomskii
superexchange \cite{Kug82,Fei97}, as well as from Jahn-Teller
distortions \cite{Zaa93}. Typically both these terms contribute jointly
to the orbital exchange interactions $J_{\gamma}$ in Eq. (\ref{Horb}),
as in LaMnO$_3$ \cite{Fei99}. Another example is the phonon-mediated
orbital exchange in spinels \cite{Rad05} which has the form of
interaction with an effective exchange $J_{\gamma}=g^2/k_{F_{1g}}$,
where $g$ is a Jahn-Teller coupling constant and $k_{F_{1g}}$ is the
elastic constant of $F_{1g}$ phonons.

For the zigzag chain of $N$ sites (assumed here to be even, and $N'=N/2$
is the number of two-site unit cells) shown in Fig. \ref{fig:chain},
the interactions with exchange constants $J_{\rm e}$ and $J_{\rm o}$
alternate between even and odd bonds, as in Eq. (\ref{gJ}),
\begin{eqnarray}
{\cal H}_{e_g}&\!=\!& H_J+H_h \nonumber \\
&\!=\!& \sum_{i=1}^{N/2}\left\{ J_{\rm o}
\left(\frac{\sqrt{3}}{2}\sigma_{2i-1}^x+\frac12\sigma_{2i-1}^y\right)\!
\left(\frac{\sqrt{3}}{2}\sigma_{2i}^x+\frac12\sigma_{2i}^y\right)
\right.\nonumber \\
& \!&+ \left. J_{\rm e}
\left(\frac{\sqrt{3}}{2}\sigma_{2i }^x -\frac12\sigma_{2i}^y \right)\!
\left(\frac{\sqrt{3}}{2}\sigma_{2i+1}^x-\frac12\sigma_{2i+1}^y\right)
\right\}
\nonumber \\
&+& \frac{h}{2}\;\sum_{i }\left(\sigma_{2i-1}^z+\sigma_{2i}^z\right).
\label{Hamiltonian1}
\end{eqnarray}
The model Eq. (\ref{Hamiltonian1}) includes a crystal field term,
\begin{equation}
H_h=\frac{h}{2}\;\sum_{i }\left(\sigma_{2i-1}^z+\sigma_{2i}^z\right),
\label{Hh}
\end{equation}
which is the source of the orbital polarization field $h$ along the
$z$-th pseudospin component. It follows from the uniform expansion or
compression of the lattice along the $c$ axis, i.e., orthogonal to the
$ab$ plane of the chain.
Although we consider for clarity $J_{\rm o}>0$ and $J_{\rm e}>0$ below,
the model is invariant with respect to the gauge transformation changing
signs of both couplings $\{J_{\rm o},J_{\rm e}\}$ simultaneously,
as alternating orbital and ferro-orbital systems are related to one
another. This can be realized explicitly by introducing the operator
${\cal U}=\Pi_{i}\sigma^z_{2i-1}$.

The Hamiltonian (\ref{Hamiltonian1}) can be exactly diagonalized
following the standard procedure for 1D systems. The Jordan-Wigner
transformation maps explicitly between pseudospin operators and spinless
fermion operators by
\begin{eqnarray}
\sigma _{j}^{+}& =&\exp\left[ i\pi\sum_{i=1}^{j-1}c_{i}^{\dagger }c_{i}^{}
\right] c_{j}^{}=\prod_{i=1}^{j-1}\sigma _{i}^{z}c_{j}^{},    \\
\sigma _{j}^{-}& =&\exp\left[-i\pi\sum_{i=1}^{j-1}c_{i}^{\dagger }c_{i}^{}
\right] c_{j}^{\dagger}=\prod_{i=1}^{j-1}\sigma_{i}^{z}c_j^{\dagger }, \\
\sigma _{j}^{z}& =&1-2c_{j}^{\dagger }c_{j}^{}.
\end{eqnarray}
Next discrete Fourier transformation for odd/even spin sites is introduced
as follows ($j=1,\dots,N'$),
\begin{eqnarray}
c_{2j-1}&=&\frac{1}{\sqrt{N'}}\sum_{k}e^{-ik j}a_{k}, \\
c_{2j}&=&\frac{1}{\sqrt{N'}}\sum_{k}e^{-ik j}b_{k},
\end{eqnarray}
with the discrete momenta $k$ which correspond to the reduced Brillouin zone
and are given by
\begin{equation}
k=\frac{n\pi}{ N^\prime  }, \quad n= -(N^\prime-1), -(N^\prime-3),
\ldots,(N^\prime -1).
\label{kset}
\end{equation}

The Hamiltonian (\ref{Hamiltonian1}) in the momentum representation becomes
a quadratic form, with mixed $k$ and $-k$ fermionic states,
\begin{eqnarray}
{\cal H}_{e_g}&=&\sum_{k} \left[ B_k^{} a_{k}^{\dagger} b_{-k}^{\dagger}
+ A_k^{}a_{k}^{\dagger} b_{k}^{}+ A_k^* b_{k}^{\dagger}a_{k}^{}
+B_k^* b_{-k}^{}a_{k}^{}\right]\nonumber \\
&+& h\sum_k(a_k^{\dagger}a_k^{} + b_k^{\dagger} b_k^{})-hN^\prime,
\label{Hamiltonian5}
\end{eqnarray}
where
\begin{eqnarray}
\label{ak}
A_k&=& J_{\rm o}+ J_{\rm e} e^{ik}, \\
\label{bk}
B_k&=& J_{\rm o} e^{i\pi/3}-J_{\rm e} e^{i(k-\pi/3)}.
\end{eqnarray}
The present Hamiltonian may be easily diagonalized by a
Bogoliubov transformation, as shown below.

\subsection{Exact solution and energy spectrum}
\label{sec:dia}

To diagonalize the Hamiltonian Eq. (\ref{Hamiltonian5}), we first
rewrite it in the symmetrized matrix form with respect to the
$k\leftrightarrow -k$ transformation,
\begin{widetext}
\begin{equation}
{\cal H}_{e_g}=\frac12\sum_{k}\;
(a_{k}^{\dagger},a_{-k}^{},b_k^{\dagger},b_{-k}^{})
\left(\begin{array}{cccc}
       h & 0  & A_k & -(P_k+Q_k)  \\
       0 & -h & -(P_k-Q_k) & -A_k \\
       A_k^*  & -(P_k^*-Q_k^*) & h & 0 \\
       -(P_k^*+Q_k^*) & -A_k^* & 0 & -h
              \end{array}\right)
\!\left( \begin{array}{c}
         a_{k} \\
         a_{-k}^{\dagger} \\
         b_k \\
         b_{-k}^{\dagger}
       \end{array}
\right),
\label{FT2}
\end{equation}
\end{widetext}
where we have introduced
\begin{eqnarray}
\label{eq:p}
P_k &\equiv&    \cos \frac{\pi}{3}\, (J_{\rm e} e^{ik}-J_{\rm o}), \\
\label{eq:q}
Q_k &\equiv& -i \sin \frac{\pi}{3}\, (J_{\rm e} e^{ik}+J_{\rm o}).
\end{eqnarray}
Eq. (\ref{FT2}) is now diagonalized by a $(4\times 4)$ Bogoliubov
transformation which connects original
$\{a_k^{\dagger},a_{-k}^{\dagger},b_k^{\dagger},b_{-k}^{\dagger}\}$
fermions with new
$\{\alpha_k^{\dagger},\alpha_{-k}^{\dagger},\beta_k^{\dagger},\beta_{-k}^{\dagger}\}$
quasiparticle (QP) operators,
\begin{eqnarray}
\left(
\begin{array}{c}
\alpha_k^{\dagger} \\
\alpha_{-k}^{} \\
\beta_k^{\dagger}\\
\beta_{-k}^{}
\end{array}
\right)=\hat{U}_{k} \left(
\begin{array}{c}
a_k^{\dagger}  \\
a_{-k}   \\
b_k^{\dagger}   \\
b_{-k}
\end{array}
\right),
\label{eq:2DXXZ_RDM}
\end{eqnarray}
where the rows of the $4\times 4$ matrix $\hat{U}_{k}$ are eigenvectors
following from:
\begin{eqnarray}
\left[{\cal H}_{e_g},\alpha_k^{\dagger}\right] &=&
\varepsilon_{k,1}\alpha_k^{\dagger}, \\
\left[{\cal H}_{e_g}, \beta_k^{\dagger}\right] &=&
\varepsilon_{k,2} \beta_k^{\dagger}.
\end{eqnarray}
Here $\varepsilon_{k,1}$ and $\varepsilon_{k,2}$ are positive energies
of elementary excitations.
After diagonalization one finds a symmetric spectrum with respect
to energy $\omega=0$, with the energies $\{\pm\varepsilon_{k,n}\}$
($n=1,2$), given by the following expressions:
\begin{eqnarray}
\label{xi}
\varepsilon_{k,1}&=&\sqrt{ C_k - \sqrt{D_k}},  \\
\label{zeta}
\varepsilon_{k,2}&=&\sqrt{ C_k + \sqrt{D_k}}.
\end{eqnarray}
This compact notation is obtained after introducing the following
definitions:
\begin{eqnarray}
\label{c}
C_k&=&\vert A_k \vert^2 + \vert P_k \vert^2+ \vert Q_k \vert^2+h^2, \\
\label{d}
D_k&=&(A_k^* P_k + A_k P_k^*)^2 -(A_k^* Q_k - A_k Q_k^*)^2 \nonumber \\
&+&(P_k^* Q_k + P_k Q_k^*)^2 + 4 \vert A_k\vert^2 h^2.
\end{eqnarray}
The obtained energies $\varepsilon_{k,1}$ (\ref{xi}) and
$\varepsilon_{k,2}$ (\ref{zeta}) are a typical result for a chain with
a unit cell consisting of two atoms. The diagonalized Hamiltonian
describes the full energy spectrum in terms of these excitations,
\begin{eqnarray}
{\cal H}_{e_g}\!=\!\sum_{k} \left\{
 \varepsilon_{k,1}\left(\alpha_k^{\dagger}\alpha_k^{}-\frac12\right)
+\varepsilon_{k,2}\left( \beta_k^{\dagger} \beta_k^{}-\frac12\right)\!\right\}.
\label{diagonalform}
\end{eqnarray}
The QP energies $\{\varepsilon_{k,1},\varepsilon_{k,2}\}$
define the excited states and give the ground state energy when
QPs are absent, similar as in the 1D QCM \cite{Brz07},
\begin{eqnarray}
\label{E0}
E_0 = -\frac12\sum_k\left(\varepsilon_{k,1}+\varepsilon_{k,2}\right).
\end{eqnarray}

In our case the chemical potential $\mu=0$ and the two bands,
$\{-\varepsilon_{k,n}\}$ ($n=1,2$), with negative energies are occupied.
In general there is an excitation gap
\begin{equation}
\label{gapdelta}
\Delta=\min_{k}\varepsilon_{k,1},
\end{equation}
and the lowest energy excitation has the energy $\Delta$. It is found
at $k=0$ and vanishes for $C_0^2=D_0$, i.e., the gap opens at the
critical field,
\begin{eqnarray}
h_c = \pm \sqrt{\vert A_0\vert^2 -\vert P_0\vert^2 -\vert Q_0\vert^2 }
= \pm\sqrt{J_{\rm o} J_{\rm e}}.
\label{conditionforcriticalh}
\end{eqnarray}
Finite $h_c$ indicates that the interactions align orbitals
perpendicular to the field in the ordered phase when $h\to 0$ and they
gradually turn at $h\to h_c$. The orbitals are aligned by the external
field in the ground state of the 60$^\circ$ compass model when $h>h_c$,
which is oriented along the $z$ direction, see Eq.
(\ref{Hamiltonian1}). We note that the ordered phase found here at
$h=0$ is in contrast to the 1D 90$^\circ$ compass model with alternating
XX and YY interactions along the zigzag chain, where the ground state is
disordered \cite{Brz07,You12b}, see also Sec. \ref{sec:qpt}.

\section{Generalized compass model}
\label{sec:gen}

\subsection{The model and exact solution}
\label{sec:model}

In the EOM Eq. (\ref{Hamiltonian1}) the interactions are fixed by the
orbital shape. For $t_{2g}$ orbitals other interactions would arise as
then the orbital flavor is conserved and the superexchange is
Ising-like \cite{Dag08,Wro10}. Such interactions resemble those in the
compass models \cite{Cin10,vdB13}, and we investigate this case below
taking the superexchange given by Eq. (\ref{gJ}).
The maximally frustrated interactions (obtained at $\theta=\pi/2$) give
the QCM and are isomorphic with the $t_{2g}$ orbital interactions
between $\{yz,zx\}$ orbitals along the zigzag chain \cite{Dag08,Wro10}.
Similar interactions are also realized between $p$ orbitals in optical
lattices \cite{Zhao08,Wu08,Gsun}, or in hyperoxides \cite{WDO11}.

The 1D GCM with $x$-th and $y$-th orbital component interactions that
alternate on even/odd exchange bonds obtained in this way is strongly
frustrated, and we study it again in finite polarization field $h$
which corresponds to a transverse magnetic field in spin systems,
\begin{eqnarray}
{\cal H}_{\rm GCM}&=& \sum_i\left\{
 J_{\rm o}\tilde{\sigma}_{2i-1}( \theta)\tilde{\sigma}_{2i  }( \theta)
+J_{\rm e}\tilde{\sigma}_{2i  }(-\theta)\tilde{\sigma}_{2i+1}(-\theta)\right\}
\nonumber \\
&-& \frac{h}{2}\;\sum_i(\sigma_{2i-1}^z+\sigma_{2i}^z).
\label{gcom}
\end{eqnarray}
At angle $\theta=\pi/3$ the EOM Eq. (\ref{Hamiltonian1}) analyzed in
Sec. \ref{sec:como} is reproduced.
Below we address a question whether the $60^\circ$ difference between
interactions along odd and even bonds in $(\sigma_x,\sigma_y)$ plane
in the EOM diminishes the short-range order induced by stronger
interactions $\propto \sigma_i^x\sigma_{i+1}^x$ along the chain.
For the numerical analysis we take $J_{\rm o}\equiv 1$
as the energy unit.

The model Eq. (\ref{gcom}) reduces to the 1D Ising model in transverse
field for $\theta=0$, and may describe the ferromagnet CoNb$_2$O$_6$,
where magnetic Co$^{2+}$ ions are arranged into near-isolated zigzag
chains along the $c$ axis with strong easy axis anisotropy due to
transverse field effects which stem from the distorted CoO$_6$ local
environment \cite{Coldea}. At $\theta=90^\circ$ the 1D GCM Eq.
(\ref{gcom}) gives a competition between two pseudospin components,
$\{\sigma_{i}^x,\sigma_{i}^y\}$ as in the 2D QCM. This case has
the highest possible frustration of interactions and the mixed terms
$\propto \sigma_i^x\sigma_{i+1}^y$, familiar from the EOM, are absent.
One can also write this model in the form of the QCM with rotated
pseudospin components,
\begin{eqnarray}
{\cal H}_{\rm QCM}&=& \sum_i\left\{
  J_{\rm o}\tilde{\sigma}_{2i-1}^x\tilde{\sigma}_{2i  }^x
+ J_{\rm e}\tilde{\sigma}_{2i  }^y\tilde{\sigma}_{2i+1}^y\right\}
\nonumber \\
&-& \frac{h}{2}\;\sum_i(\tilde{\sigma}_{2i-1}^z+\tilde{\sigma}_{2i}^z).
\label{gcom90}
\end{eqnarray}
where the rotation by angle $\theta=\pm\pi/2$ with respect to the $z$
axis in the pseudospin space is made on even/odd bonds \cite{Cin10}.

In two dimensions the Ising-like order is determined by the strongest
interaction $\propto\sigma_m^x\sigma_n^x$ as long as $\theta<\theta_c$
\cite{Cin10}, and the mixed interactions $\propto\sigma_m^x\sigma_n^z$
play no role in this regime. Existence of a second-order QPT from the
Ising order to the compass-like nematic order was established at
$\theta_c=84.8^\circ$ using the multiscale entanglement renormalization
ansatz (MERA) \cite{Cin10}. Here we explore ground states of the 1D QCM
Eq. (\ref{gcom}) in the entire parameter space and investigate whether
signatures of a similar transition may be recognized in the
thermodynamic quantities, the susceptibility and the specific heat.

The GCM Eq. (\ref{gcom}) can be solved exactly following the
same steps as described in Sec. \ref{sec:como}, and this solution is
equivalent at angle $\theta=90^\circ$ to that given in Ref.
\onlinecite{Brz07}. We introduce $A_k$ defined by Eq. (\ref{ak}), and
\begin{eqnarray}
P_k^{'}&\equiv&(J_{\rm e} e^{ik}-J_{\rm o})\cos\theta\,, \\
Q_k^{'}&\equiv& -i(J_{\rm e} e^{ik}+J_{\rm o})\sin\theta\,,
\end{eqnarray}
which reproduce Eqs. (\ref{eq:p}) and (\ref{eq:q}) at $\theta=\pi/3$.
The algebraic structure of the exact solution is now the same as in Sec.
\ref{sec:eg}, and the excitation energies $\varepsilon_{k,1}$ and
$\varepsilon_{k,2}$ are given by Eqs. (\ref{xi}) and (\ref{zeta}), with
\begin{eqnarray}
C_k^{'}&=&\vert A_k\vert^2 + \vert P_k^{'}\vert^2+\vert Q_k\vert^2+h^2,
\nonumber\\
&=&2J_{\rm o}^2+2J_{\rm e}^2+4 \sin^2\theta J_{\rm o}J_{\rm e}\cos k+h^2,
\\
D_k^{'}&=&[A_k^*P_k^{'}+A_k(P_k^{'})^*]^2-[A_k^*Q_k^{'}-A_k(Q_k^{'})^*]^2
\nonumber \\
&+&[(P_k^{'})^*Q_k^{'} + P_k^{'}(Q_k^{'})^*]^2 +4 \vert A_k\vert^2 h^2,
\end{eqnarray}
which replace now $C_k$ and $D_k$ given by Eqs. (\ref{c}) and (\ref{d})
for the EOM. We note that the negative QP energies,
$-\varepsilon_{k,n}$ for $n=1,2$, correspond to the filled bands in the
fermionic representation. They serve to evaluate the ground state energy
for the GCM, and one may use again Eq. (\ref{E0}). Actually, the
convention used here sets this energy at the energy origin, and
therefore the free energy considered in Sec. \ref{sec:qpt} starts from
zero at $T=0$.

\begin{figure}[t!]
\includegraphics[width=7.5cm]{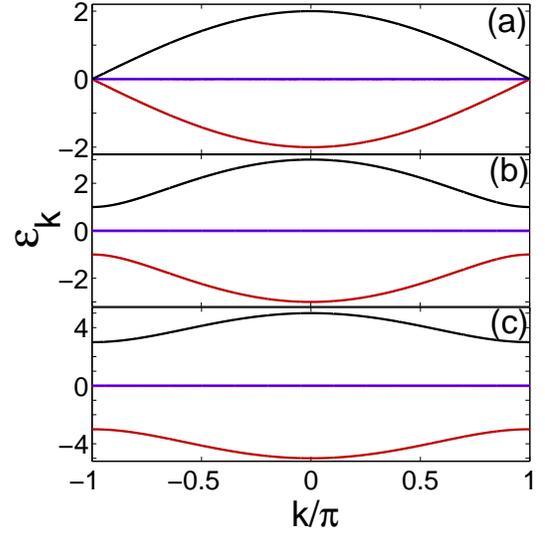}
\caption{(Color online)
The electronic QP energies $\pm\varepsilon_{k,n}/2$ ($n=1,2$) as
obtained for the QCM Eq. (\ref{gcom90}) with increasing values of $J_{\rm e}$:
(a) $J_{\rm e}=1$,
(b) $J_{\rm e}=2$, and
(c) $J_{\rm e}=4$.
Except for the isotropic case (a) of $J_{\rm e}=J_{\rm o}$, the spectra are
characterized by a finite gap between $\varepsilon_{k,n}$ and
$\varepsilon_{k,2}$.
Parameters: $J_{\rm o}=1$, $h=0$, and $\theta=\pi/2$.}
\label{fig:allek}
\end{figure}

The case of angle $\theta=\pi/2$ in the 1D GCM is special and will
be considered in more detail below. The structure of the Hilbert
space gives here a macroscopic degeneracy of $2^{N/2-1}$ away from the
isotropic point, and the enhanced degeneracy of $2^{N/2}$ when the orbital
interactions are isotropic, i.e., $J_{\rm e}=J_{\rm o}$. We recall that
we use here odd numbers of $k$ values included in the chosen set
given by Eq. (\ref{kset}), and only in the thermodynamic limit
we recover the degeneracy of $2\times 2^{N/2}$ for isotropic
interactions \cite{Brz07}. Using fermions after the Jordan-Wigner
transformation, this degeneracy is due to the acoustic branch which has
no dispersion and is found at zero energy, $\varepsilon_{k,1}=0$, see
Fig. \ref{fig:allek}. Then this branch is half-filled by fermions as it
becomes degenerate with the one of negative energy $-\varepsilon_{k,1}$.
Therefore, using the fermionic language one recovers here a macroscopic
$2^{N/2}$ degeneracy of the ground state in the thermodynamic limit,
independently of the mutual values of exchange parameters, and one finds
for $J_{\rm e}\neq J_{\rm o}$ that
$\forall k$: $\varepsilon_{k,1}<\varepsilon_{k,2}$,
see Figs. \ref{fig:allek}(b) and \ref{fig:allek}(c). The gap at $k=\pi$
is given by the anisotropy of the pseudospin exchange,
$\Delta=|J_{\rm e}-J_{\rm o}|$. The situation changes, however, when
$J_{\rm e}=J_{\rm o}$ and the gap between $\varepsilon_{\pi,2}$ and
$\varepsilon_{\pi,1}$ closes, see Fig. \ref{fig:allek}(a).
This implies that the degeneracy increases
by an additional factor of 2 due to the band-edge points.

\begin{figure}[t!]
\includegraphics[width=8cm]{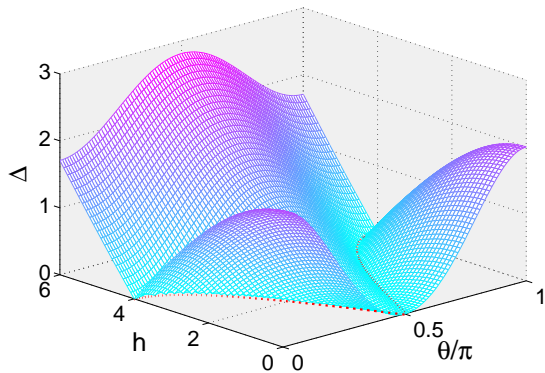}
\caption{(Color online) The gap $\Delta$ as a function of $\theta$ and
$h$. The dotted line is the critical line given by Eq.
(\ref{conditionforcriticalh2}).
Parameters: $J_{\rm o}=1$, $J_{\rm e}=4$.}
\label{fig:gap}
\end{figure}

\begin{figure}[b!]
\includegraphics[width=7.5cm]{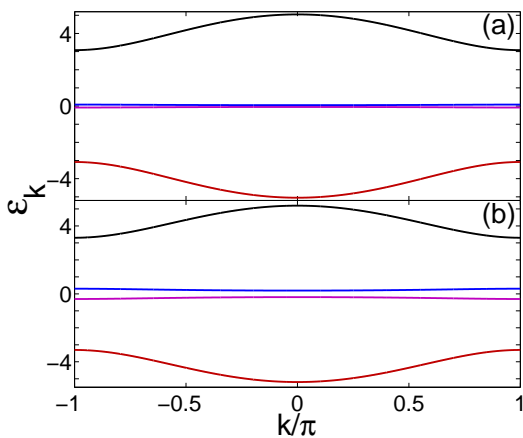}
\caption{(Color online)
The electronic QP energies $\pm\varepsilon_{k,n}/2$ ($n=1,2$) as obtained
for the QCM Eq. (\ref{gcom90}) at finite polarization field (\ref{Hh}):
(a) $h=1$, and
(b) $h=2$.
In both cases the lower two bands are filled by fermions and a finite
gap separates occupied from empty bands.
Parameters: $J_{\rm o}=1$, $J_{\rm e}=4$, and $\theta=\pi/2$. }
\label{fig:ek_h}
\end{figure}

High degeneracy of the ground state is removed by finite field $h>0$.
For $\theta\neq\pi/2$, Eq. (\ref{conditionforcriticalh}) reduces
to,
\begin{eqnarray}
h_c = \pm 2\cos\theta\sqrt{J_{\rm o} J_{\rm e}}.
\label{conditionforcriticalh2}
\end{eqnarray}
It defines the critical field at which the gap closes, see
Fig. \ref{fig:gap}.

As $h$ approaches $h_c$, the gap vanishes as $\Delta\sim(h-h_c)^{\nu z}$,
where $\nu$ and $z$ are the correlation-length and dynamic exponent,
respectively. The gap near criticality is
\begin{equation}
\label{Delta}
\Delta \simeq \frac{h^2-h_c^2}{2(J_{\rm o}+J_{\rm e})},
\end{equation}
and one finds the critical exponent $\nu z=1$. In this sense, the 1D QCM
has Ising-type long range order for finite $\theta<\pi/2$ and $h<h_c$.
This is in analogy to the Ising model in transverse magnetic field,
where a similar transition was reported \cite{Dzi05}.
We emphasize that the phase space of the orbital liquid consists thus of
a plane in the parameter space, spanned by $\{J_{\rm e},J_{\rm o}\}$.

The critical lines intersect at $\theta_c=\pi/2$ and $h_c=0$, forming
a multicritical point, where the model is gapless irrespective of the
values of $J_{\rm e}$ and $J_{\rm o}$, see Fig. \ref{fig:gap}. It has
been proven that the 90$^\circ$ quantum compass model is critical for
arbitrary ratio $J_{\rm e}/J_{\rm o}$ and the point
$J_{\rm e}=J_{\rm o}$ corresponds to a multicritical point \cite{Eri09}.
Finite field $h$ polarizes orbitals and removes high degeneracy of
the ground state. For the fermionic QP bands this means that
a gap at the Fermi energy opens exponentially between the bands
$\varepsilon_{k,1}$ and $-\varepsilon_{k,1}$, and the system turns into
an insulator, see Fig. \ref{fig:ek_h}. The gap is much smaller than the
field $h$ and therefore the thermal excitations through the gap
contribute to the thermodynamic properties at relatively low temperature
as we show below in Sec. \ref{sec:qpt}.

\subsection{Orbital order and correlation functions at finite field}
\label{sec:corr}

\begin{figure}[t!]
\includegraphics[width=8cm]{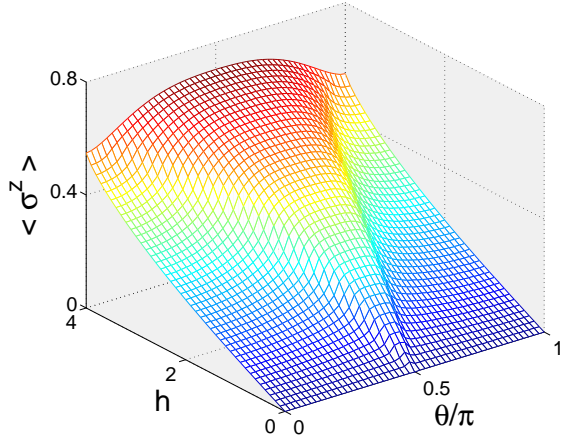}
\caption{(Color online) Orbital polarization
$\langle\sigma^z\rangle$ obtained for the GCM with increasing
field $h$ in the ground state for different values of $\theta$.
Parameters: $J_{\rm o}=1$, $J_{\rm e}=4$.}
\label{fig:sz}
\end{figure}

Frustrated interactions in Eq. (\ref{gcom}) result in disordered state
and the longitudinal polarization vanishes at $T=0$, i.e.,
$\langle\sigma_{i}^{x}\rangle=\langle \sigma_{i}^{y}\rangle=0$.
The transverse polarization,
\begin{equation}
\label{m}
{\cal P}=N\langle\sigma_{i}^{z}\rangle,
\end{equation}
is induced by finite field $h$ at $T=0$;
it is found with help of Hellmann-Feynman theorem,
\begin{equation}
\label{sz}
{\cal P}=-\frac{\partial E_0 }{\partial h}.
\end{equation}
A similar thermodynamic relation which involves the total spectrum
via the free energy ${\cal F}$ is used to determine
$\langle\sigma_i^z\rangle$ at finite $T>0$ in Sec. \ref{sec:qpt}.
The order parameter $\langle\sigma_i^z\rangle$ is induced by the
transverse field $h$, as shown in Fig. \ref{fig:sz}. By investigating
the behavior of $\langle\sigma_{i}^{z}\rangle$ with increasing field
$h$, we establish that the field-induced QPT is here second order for
any angle $\theta$ \cite{Sun09}. It is also accompanied by a scaling
behavior since the correlation length diverges and there is no
characteristic length scale in the system at the critical point.

However, one finds a qualitatively different behavior at finite field
$h$ for the GCM with interactions at $\theta<\pi/2$ (which includes the
EOM) from that at $\theta=\pi/2$ for the QCM. The disordered phase in
the QCM may easily be polarized by the field, while the ground state is
more robust away from this point. In this regime the model has
N\'eel order induced by the $x$-th pseudospin components (Ising order
for the strongest interaction) and is harder to be destroyed by the
transverse field. The results shown in Fig. \ref{fig:sz} are confirmed
by exact diagonalization that we performed on finite clusters in
addition. Increasing transverse field induces finite
$\langle\sigma_i^z\rangle$ and drives the system into a saturated
polarized phase found above the critical field, i.e., for $h>h_c$.

\begin{figure}[t!]
\includegraphics[width=8cm]{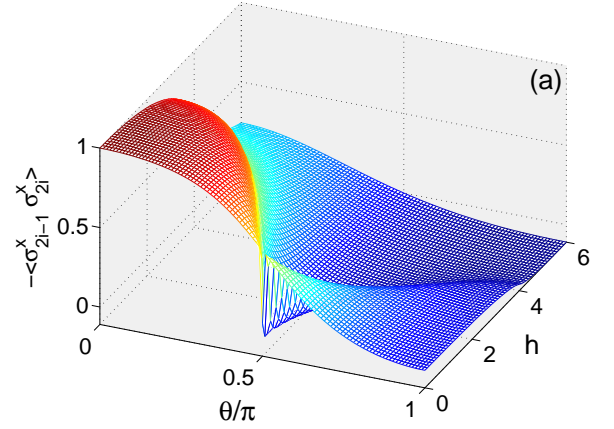}
\includegraphics[width=8cm]{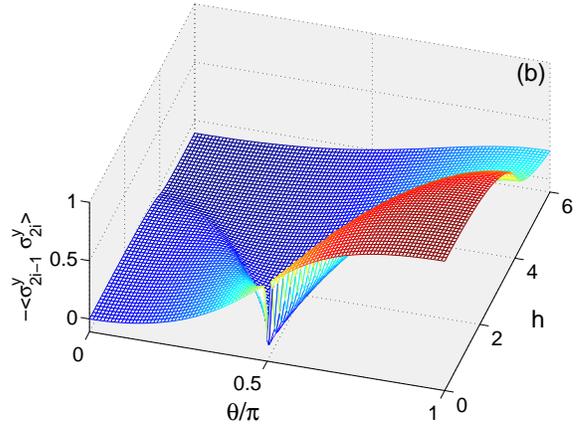}
\caption{(Color online)
The 3D panorama of the NN orbital correlations, shown as functions
of angle $\theta$ and the transverse field $h$ on odd bonds:
(a) $-\langle\sigma_{2i-1}^x\sigma_{2i}^x\rangle$, and
(b) $-\langle\sigma_{2i-1}^y\sigma_{2i}^y\rangle$.
Parameters: $J_{\rm o}=1$ and $J_{\rm e}=4$.}
\label{fig:ss}
\end{figure}

Two-point correlation functions which correspond to the dominating
interaction decay algebraically with distance $r$ \cite{Brz07}. They
are given by \cite{Osb02}:
\begin{eqnarray}
\label{sxsx}
\langle \sigma_0^x \sigma_r^x \rangle &=& \left \vert
\begin{array}{c c c c}
G_{-1} & G_{-2} & \cdot & G_{-r} \\
G_{0} & G_{-1} & \cdot & G_{-r+1} \\
\vdots & \vdots & \ddots &\vdots  \\
G_{r-2} & G_{r-3} & \cdot & G_{-1}
\end{array} \right \vert, \\
\label{sysy}
\langle \sigma_0^y\sigma_r^y \rangle &=& \left \vert
\begin{array}{c c c c}
G_{1} & G_{0} & \cdot & G_{-r+2} \\
G_{2} & G_{1} & \cdot & G_{-r+3} \\
\vdots & \vdots & \ddots &\vdots  \\
G_{r} & G_{r-1} & \cdot & G_{1}
\end{array} \right \vert, \\
\label{szsz}
\langle \sigma_0^z\sigma_r^z \rangle &=&
4\langle\sigma^z\rangle ^2 - G_r G_{-r},
\end{eqnarray}
where we have introduced the short-hand notation for the mixed
correlation function,
\begin{eqnarray}
G_r=\langle \sigma_0^y\sigma_r^x \rangle.
\end{eqnarray}

The numerical analysis shows two distinct phases at $h=0$, with large
either $-\langle\sigma_{2i-1}^x\sigma_{2i}^x\rangle$ or
$-\langle\sigma_{2i-1}^y\sigma_{2i}^y\rangle$, depending on whether
$\theta<\pi/2$ or $\theta>\pi/2$. Note that NN orbital correlations are
almost classical in a broad range of $\theta$ at $h=0$ as the model is
Ising-like. The correlations decrease, however, when the quantum
critical point (QCP) at $\theta=\pi/2$ is approached \cite{Brz07}.
At this point one finds the disordered orbital state and the role of
XX and YY correlations is interchanged, see Fig. \ref{fig:ss}.
In both phases at $\theta\neq\pi/2$ there is a gap in the excitation
spectrum which vanishes at the critical field ($h$=$h_c$), together
with a jump in transverse magnetization shown in Fig. \ref{fig:sz}
and in the NN orbital correlation functions in Fig. \ref{fig:ss} at
$h_c(\theta)$.

We remark that the vanishing of the intersite correlators between
uncoupled orbitals in the 1D QCM follows indeed from the local $Z_2$
symmetry, see the Appendix, and may also be seen as a consequence of
Elitzur's theorem --- similar as in case of the 2D Kitaev model on a
hexagonal lattice \cite{Che08}. One may also employ the general
approach of "bond algebra" \cite{Nus09} which leads to the same conclusion.

\section{Finite temperature properties}
\label{sec:qpt}

\subsection{The entropy and the cooling rate}
\label{sec:s}

Having the exact solution of the GCM (\ref{gcom}), it is straightforward
to obtain its full thermodynamic properties at finite temperatures.
For the particle-hole excitation spectrum (\ref{diagonalform}),
we determined the free energy of the quantum spin chain per site
(here and below we take the Boltzmann constant $k_{\rm B}\equiv 1$),
\begin{eqnarray}
\label{free}
{\cal F}=
-T\sum_k\sum_{j=1}^2\ln\left(2\cosh\frac{\varepsilon_{k,j}}{2T}\right).
\end{eqnarray}
Entropy ${\cal S}$ provides information about the evolution of spectra
with increasing transverse field $h$. It has been determined from the
free energy ${\cal F}$ (\ref{free}) via the usual thermodynamic relation,
\begin{eqnarray}
{\cal S}&\!=&-\left(\frac{\partial{\cal F}}{\partial T}\right)_V
\nonumber  \\
&\!=&\!\sum_k\sum_{j=1}^{2}\ln\left(2\cosh\frac{\varepsilon_{k,j}}{2T}\right)
-\sum_k\sum_{j=1}^{2}\left( \frac{\varepsilon_{k,j}}{2T}
\tanh\frac{\varepsilon_{k,j}}{2T}\right). \nonumber \\
\end{eqnarray}

\begin{figure}[t!]
\includegraphics[width=8cm]{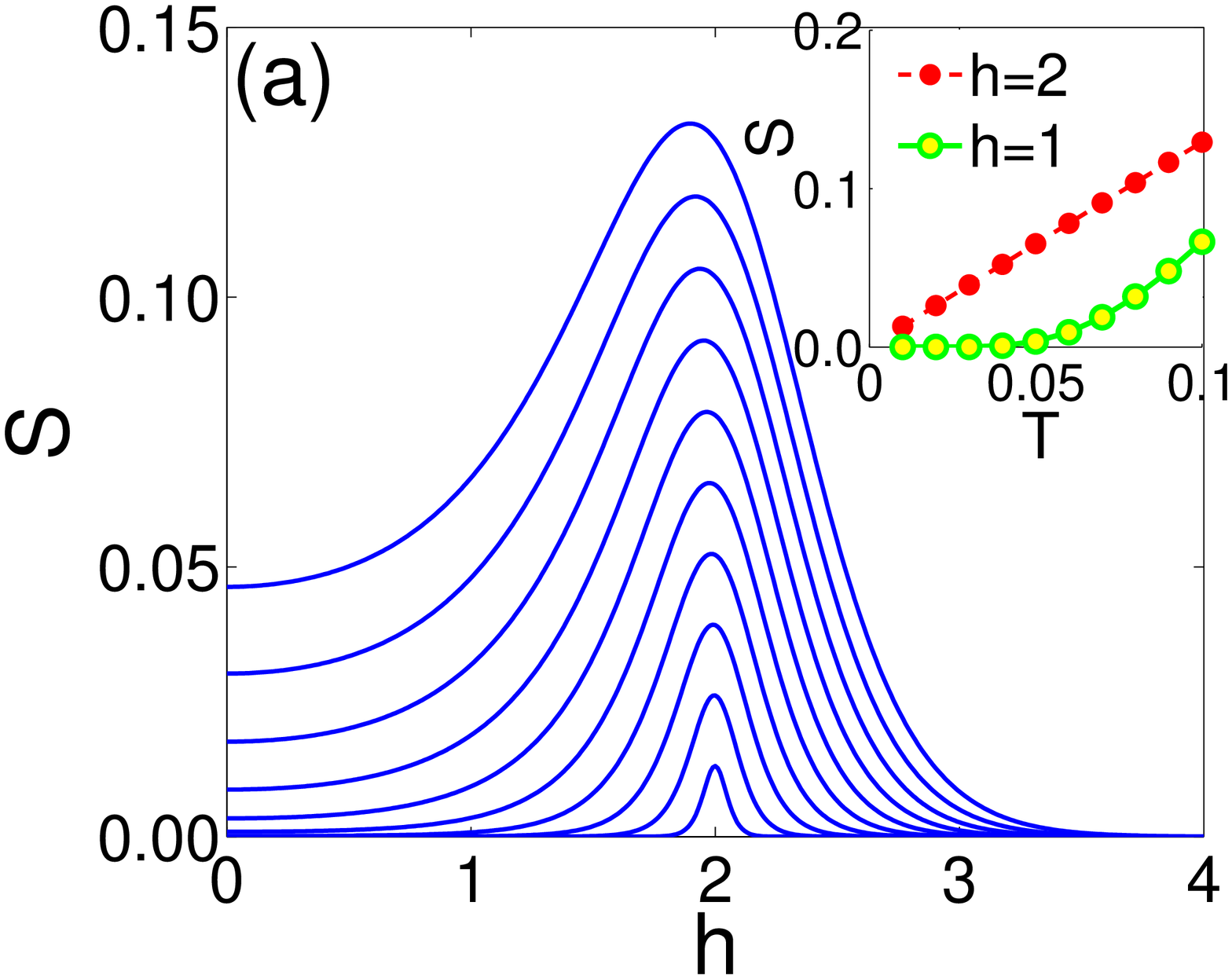}
\includegraphics[width=8cm]{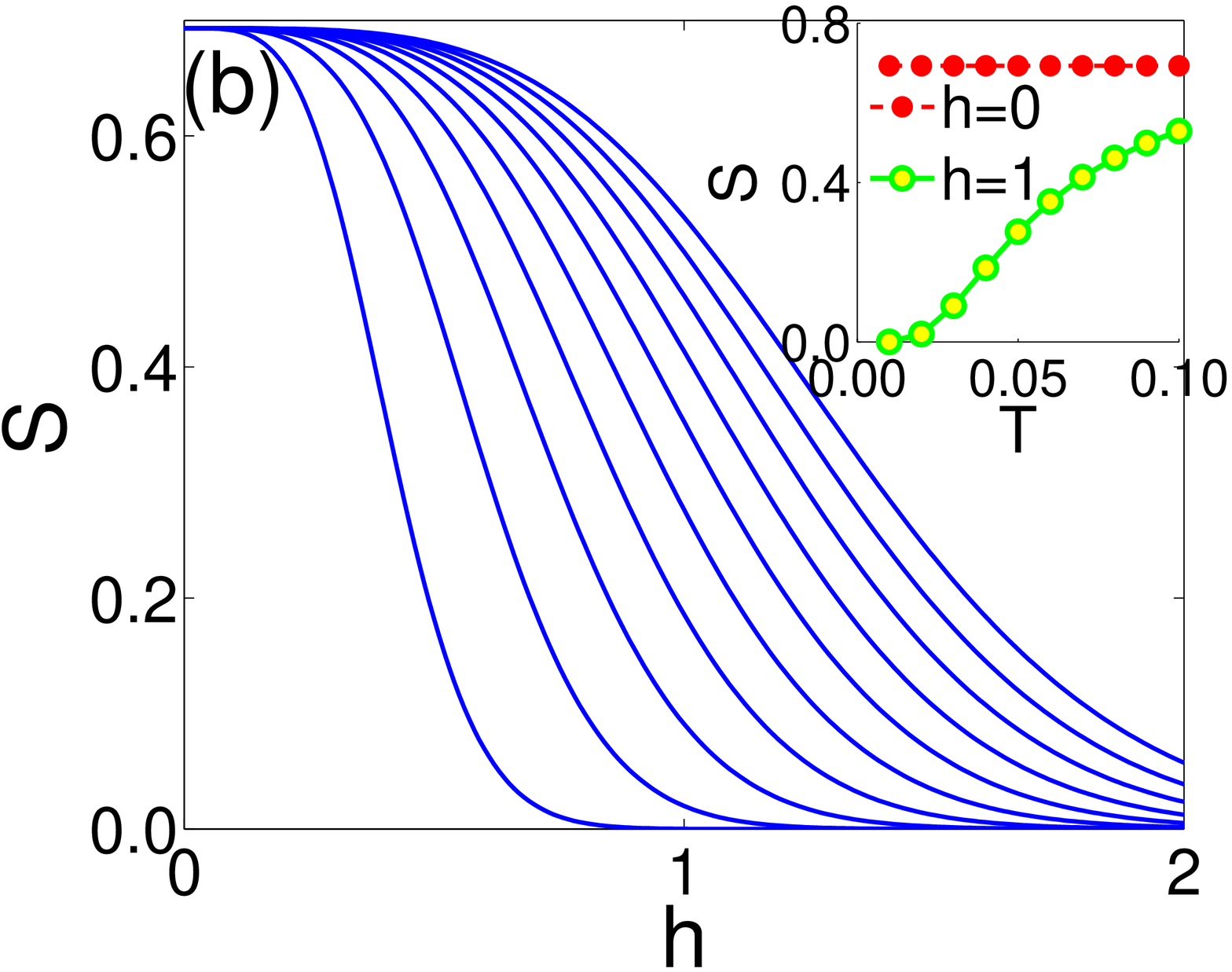}
\caption{(Color online)
The entropy ${\cal S}$ per unit cell for increasing field $h$ at
different temperature $T=0.01,0.02,\cdots 0.10$ (from bottom to top)
for two values of $\theta$:
(a) the EOM ($\theta=\pi/3$), and
(b) the QCM ($\theta=\pi/2$),
corresponding to the critical field $h_c$=2 and 0, respectively.
Insets show the temperature scaling of entropy for the critical
field (top lines) and for the noncritical case (bottom lines).
Parameters: $J_{\rm o}=1$, $J_{\rm e}=4$.}
\label{fig:ent}
\end{figure}

For the EOM, the entropy vanishes at $h=0$ and at $T=0$.
It grows with increasing $T$ when thermal excitations gradually include
more and more of excited states and this increase is faster at finite
field, for instance finite entropy is found already at $T>0.05$ if
$h=1$, see inset in Fig. \ref{fig:ent}(a). The entropy displays a
distinct maximum for increasing transverse field at $h\simeq h_c$,
where the gap closes, see Fig. \ref{fig:ent}(a), implying the QCP.
This accumulation of entropy close to the QCP indicates that the states
which characterize competing phases are almost degenerate and the
system is "maximally undecided" which ground state to choose \cite{Wu11}.
The landscape of ${\cal S}$ defines the quantum critical regime, where
$T\gg\Delta$ and role played by quantum and thermal fluctuations is
equally important for the dynamics \cite{Sac00}. Especially, the system
is gapless along the critical line and the entropy is linear in $T$,
i.e., ${\cal S}\propto T$ for low temperatures, while in the gapped
phases an exponential behavior, i.e., ${\cal S}\propto\exp(-\Delta/T)$
is observed.

\begin{figure}[t!]
\includegraphics[width=8cm]{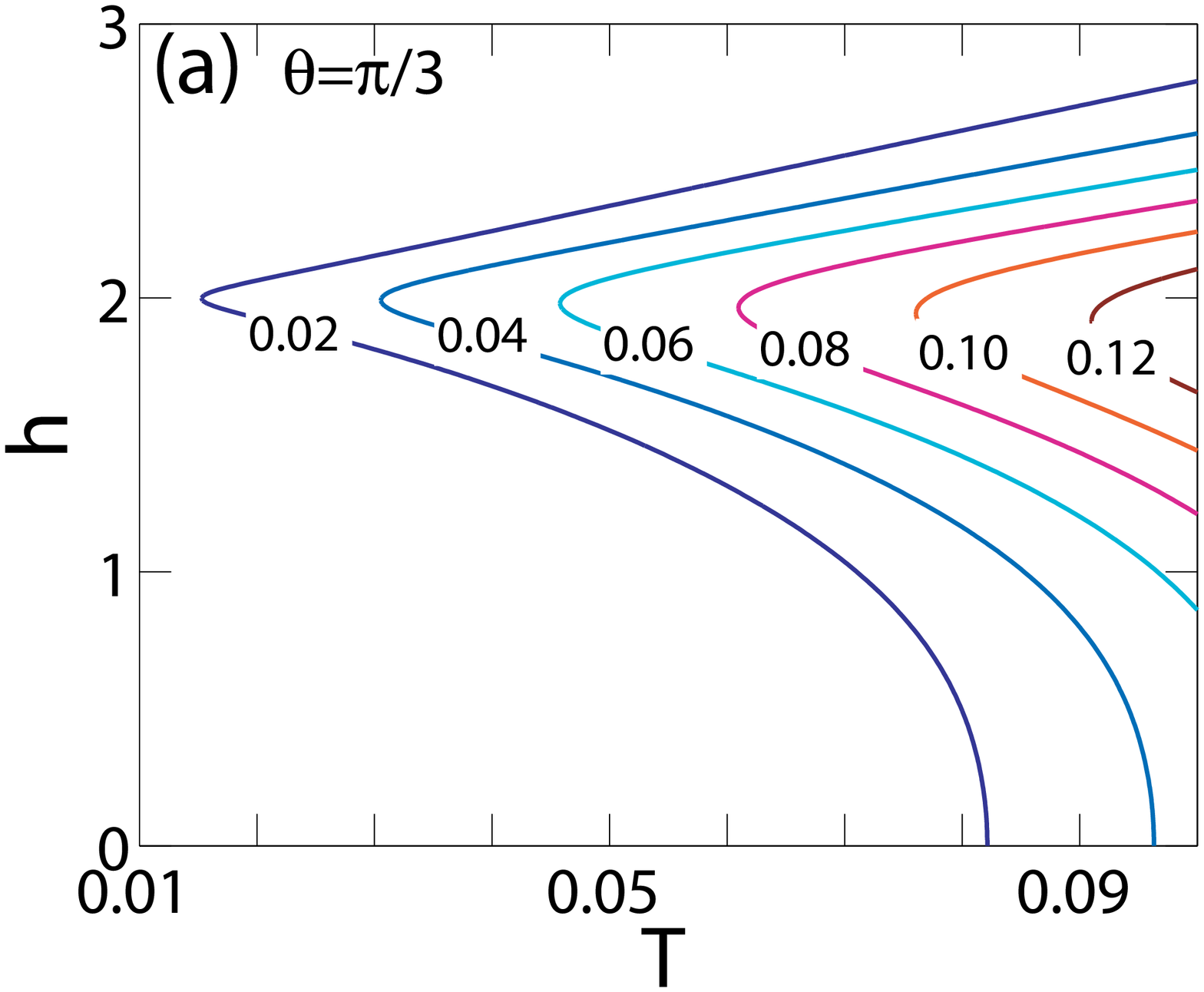}
\includegraphics[width=8cm]{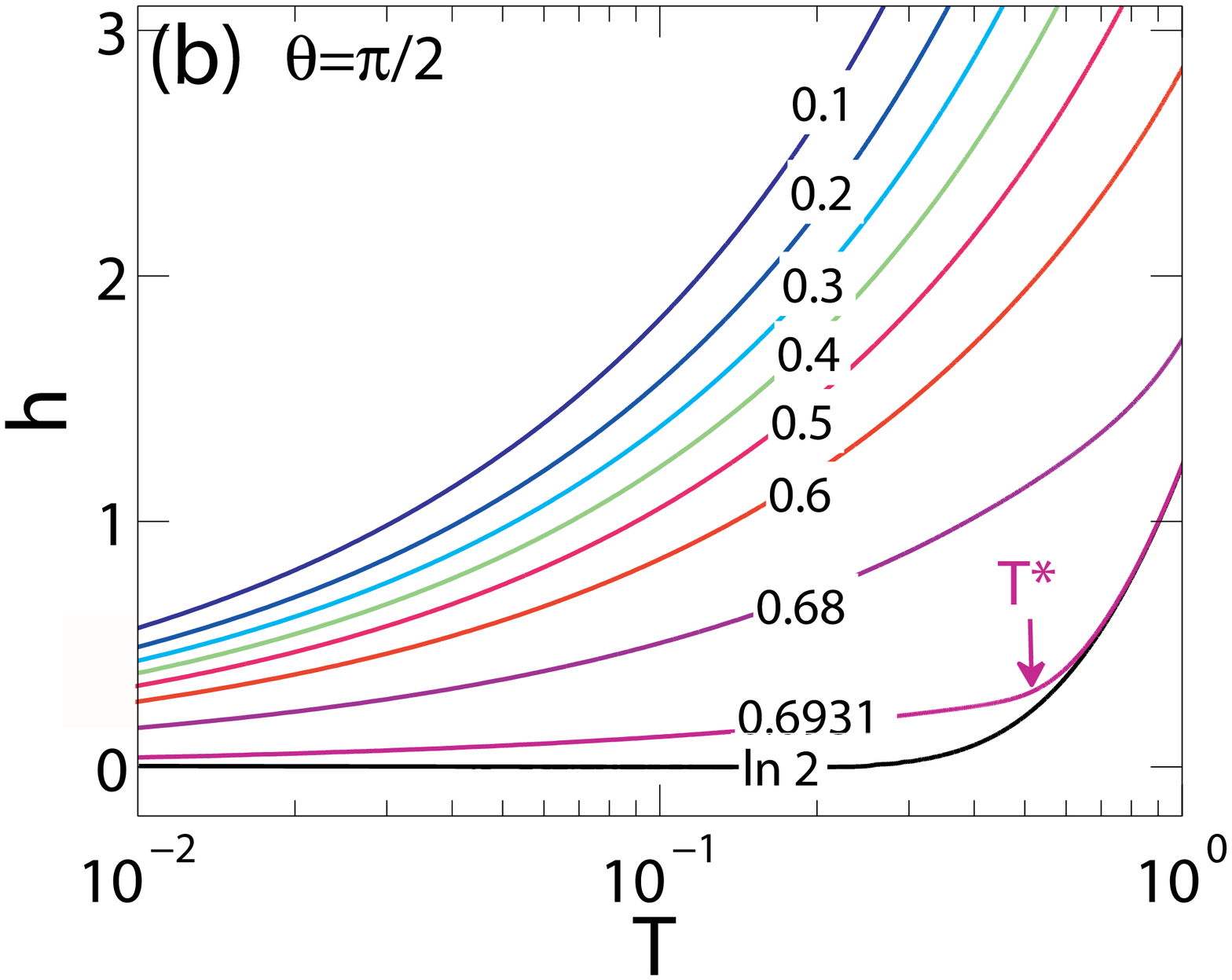}
\caption{(Color online)
Lines of constant entropy ${\cal S}$ per unit cell, i.e., adiabatic
demagnetization curves of the extended QCM in a transverse transverse
field, as obtained for:
(a) the EOM ($\theta=\pi/3$), and
(b) the QCM ($\theta=\pi/2$).
The QCP at $T=0$ gives a very distinct behavior in both cases near the
critical field, being $h_c=2$ for the EOM (a), and $h_c=0$ for the QCM
(b). In case (b) the field $h$ corresponding to a constant entropy
exhibits a logarithmic increase with temperature below $T^*$.
Parameters: $J_{\rm o}=1$, $J_{\rm e}=4$.}
\label{fig:map}
\end{figure}

In the 1D QCM one finds a different behavior, see Fig. \ref{fig:ent}(b).
The entropy ${\cal S}$  approaches here $\ln 2$ which follows from the
high degeneracy $2^{N/2-1}$ of the disordered ground state. At $h=0$
one finds here a {\it macroscopic\/} entropy ${\cal S}\simeq\ln 2$ per
unit cell that does not change with increasing temperature $T$ over a
temperature range below the crossover temperature $T^*$, see below.

The qualitative difference between the EOM and the 1D QCM is best
illustrated by the lines of constant entropy. The entropy ${\cal S}$
vanishes for the EOM at $T=0$, see Fig. \ref{fig:map}(a), where the
strongest interactions impose the quasi-order in the ground state.
This follows the third law of thermodynamics which states that for pure
and uniform phases the entropy falls to zero at $T\to 0$. However, in
the vicinity of $h_c=2$ it increases fast with increasing $T$.

In contrast, the entropy for the QCM is maximal,
${\cal S}_{\rm max}=\ln 2$, at the QCP at $h_c=0$, and finite $h$
{\it reduces\/} ${\cal S}$ rapidly. In the vicinity of the QCP the
field corresponding to a constant entropy exhibits a logarithmic
increase with temperature, $h\propto\ln T$, see Fig. \ref{fig:map}(b).
This behavior demonstrates that the high degeneracy of the ground
state is reduced by the external field which selects only certain
states with their symmetry adapted to the field. A similar reduction
of the ground state degeneracy is found in the 2D QCM when the added
Heisenberg spin couplings induce magnetic long-range order \cite{Tro10}.

The entropy in the QCM is almost insensitive to increasing temperature,
but the field quenches the spin disorder leading to a crossover to the
classical state. These features could be the subject of future
experimental studies. Recently, the complete entropic landscape was
quantitatively measured for Sr$_3$Ru$_2$O$_7$ under transverse field in
the vicinity of quantum criticality \cite{Rost09}. More interestingly,
the low-entropy state has been a grand concern in realizing some exotic
phases in optical lattice such as $d$-wave superconductivity
\cite{Ho09,McK11}.

The field-induced QPT leads to universal responses when the applied
field is varied adiabatically, and the magnetocaloric effect (MCE)
can be used to study their quantum criticality. The adiabatic
demagnetization curves of extended quantum models,
${\cal S}(h,T)={\rm const}$, are shown in Fig. \ref{fig:map}. The MCE
is closely related to the generalized cooling rate defined as follows,
\begin{equation}
\Gamma_h=-\frac{1}{T}\frac{(\partial{\cal S}/\partial h)_T}
{(\partial{\cal S}/\partial T)_h}=
\frac{1}{T}\left(\frac{\partial T}{\partial h}\right)_{\cal S}.
\label{Gammah}
\end{equation}

\begin{figure}[t!]
\includegraphics[width=8cm]{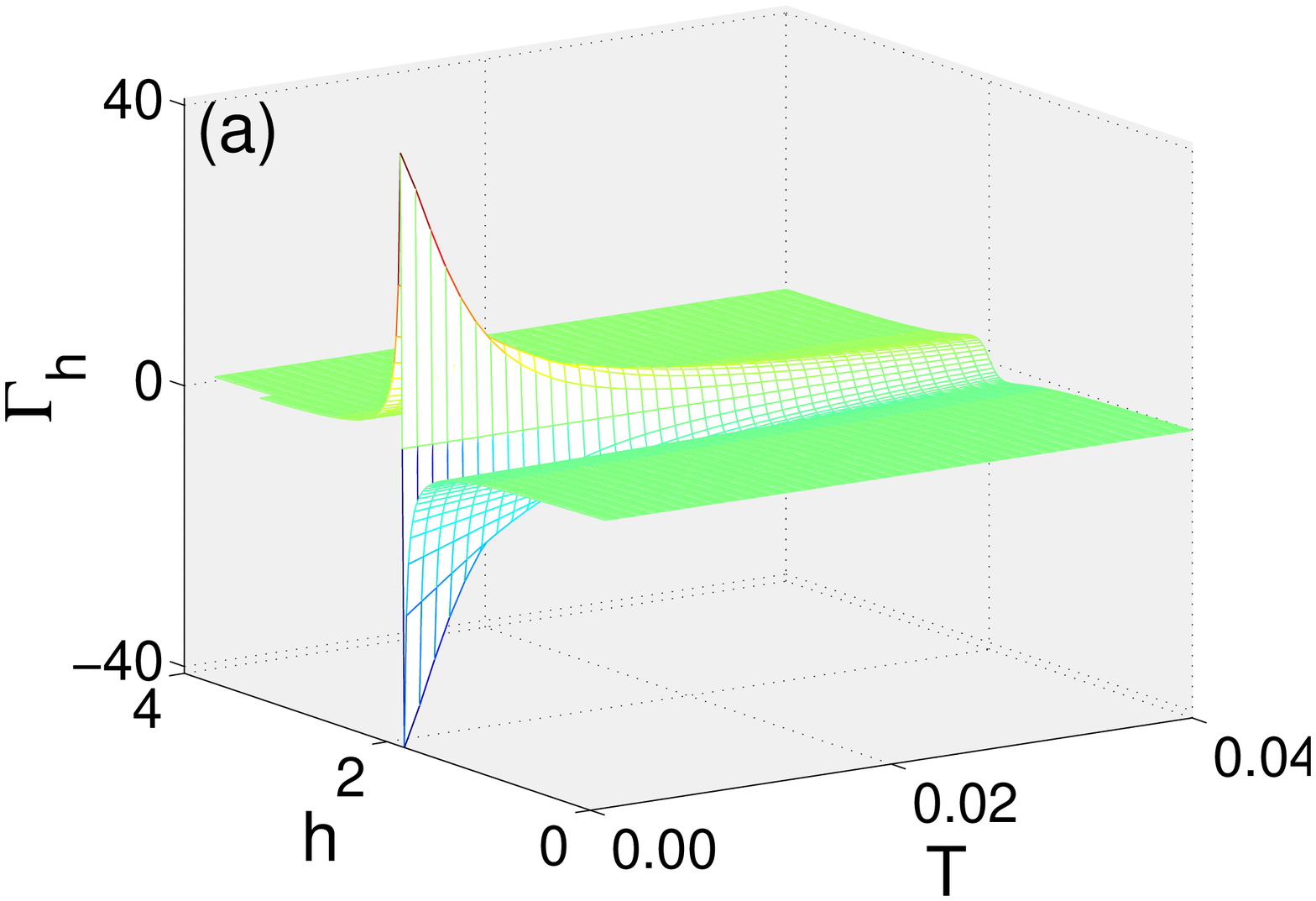}
\includegraphics[width=8cm]{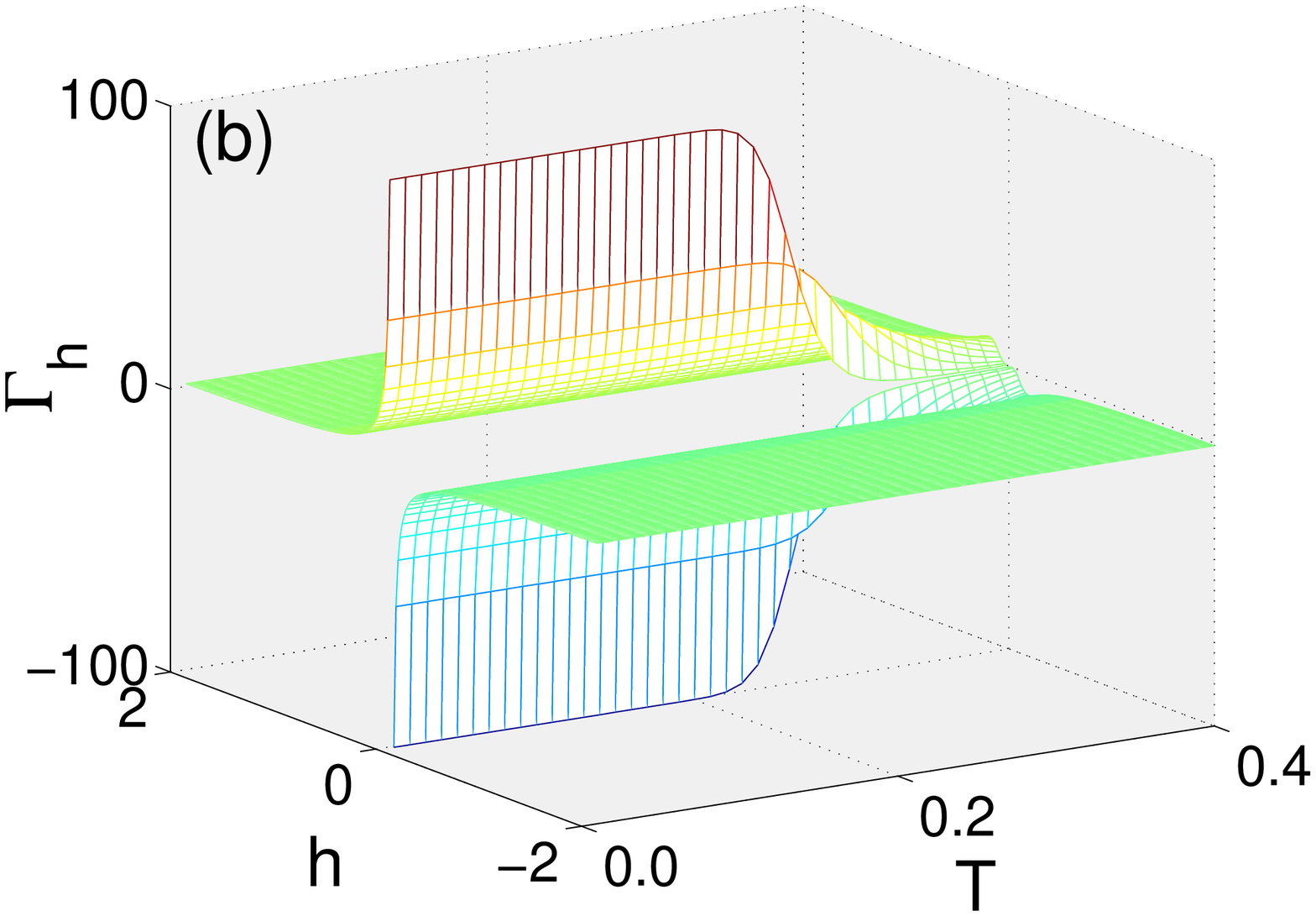}
\caption{(Color online) The cooling rate $\Gamma_h$ Eq. (\ref{Gammah})
as obtained for increasing field $h$ and temperature $T$ at:
(a) $\theta=\pi/3$, and
(b) $\theta=\pi/2$.
Parameters: $J_{\rm o}=1$, $J_{\rm e}=4$. }
\label{fig:cool}
\end{figure}

Generally, the variation of entropy ${\cal S}$ with external field $h$
is more singular than that of the specific heat considered in Sec.
\ref{sec:cv}, so one expects that the MCE Eq. (\ref{Gammah}) is
particularly large in the vicinity of the QCP. Near a field-tuned QCP,
the critical part of the free energy takes usually the hyperscaling
form in $d$ dimensions \cite{Zhu03}, $F=F_0 T^{d/z+1}f(x/T^{1/\nu z})$,
where $x=h-h_c$. The universal function $f(x)$ has diverse
asymptotic behaviors in the $x \to 0$ and $x\to\pm\infty$ limits,
respectively, corresponding to the quantum critical and quantum
disordered/renormalized classical regimes.
This divergent behavior at the QCP obeys a universal scaling law
\cite{Zhu03},
\begin{eqnarray}
\Gamma_h(T\to 0,h)=-\,G_h\,\frac{1}{h-h_c},
\end{eqnarray}
where a universal amplitude $G_h=1$ is found. This value is expected
for a $Z_2$ symmetry in one dimension. In the opposite limit,
$\Gamma_h\sim 1/T^{1/\nu z}$ for $x\ll T$. The $1/x$ divergence in the
low temperature limit amounts to a sign change of $\Gamma_h$ as
entropy accumulates near a QCP, as shown in Fig. \ref{fig:cool}(a).
Therefore, the critical fields are pinpointed by sign changes of
$\Gamma_h$ from negative to positive values upon increasing field.
As the temperature is raised, the discontinuity at $h_c$ is rapidly
reduced and all the distinct features seen at $T=0$ gradually disappear.

The dependence of the cooling rate on $h$, found for the disordered
ground state of the QCM (at $\theta=\pi/2$), is qualitatively different,
see Fig. \ref{fig:cool}(b). One finds here sharp and pronounced positive
and negative peaks which occur at the transition point $h_c=0$, and
this structure is robust, i.e., the strength of these peaks does not
vary upon increasing temperature until a critical value is reached.
The strong enhancement of the MCE arising from quantum fluctuations
near a $h$-induced QCP can be used for finding an efficient and flexible
high performance field cooling over an extended temperature range.

\begin{figure}[t!]
\includegraphics[width=8cm]{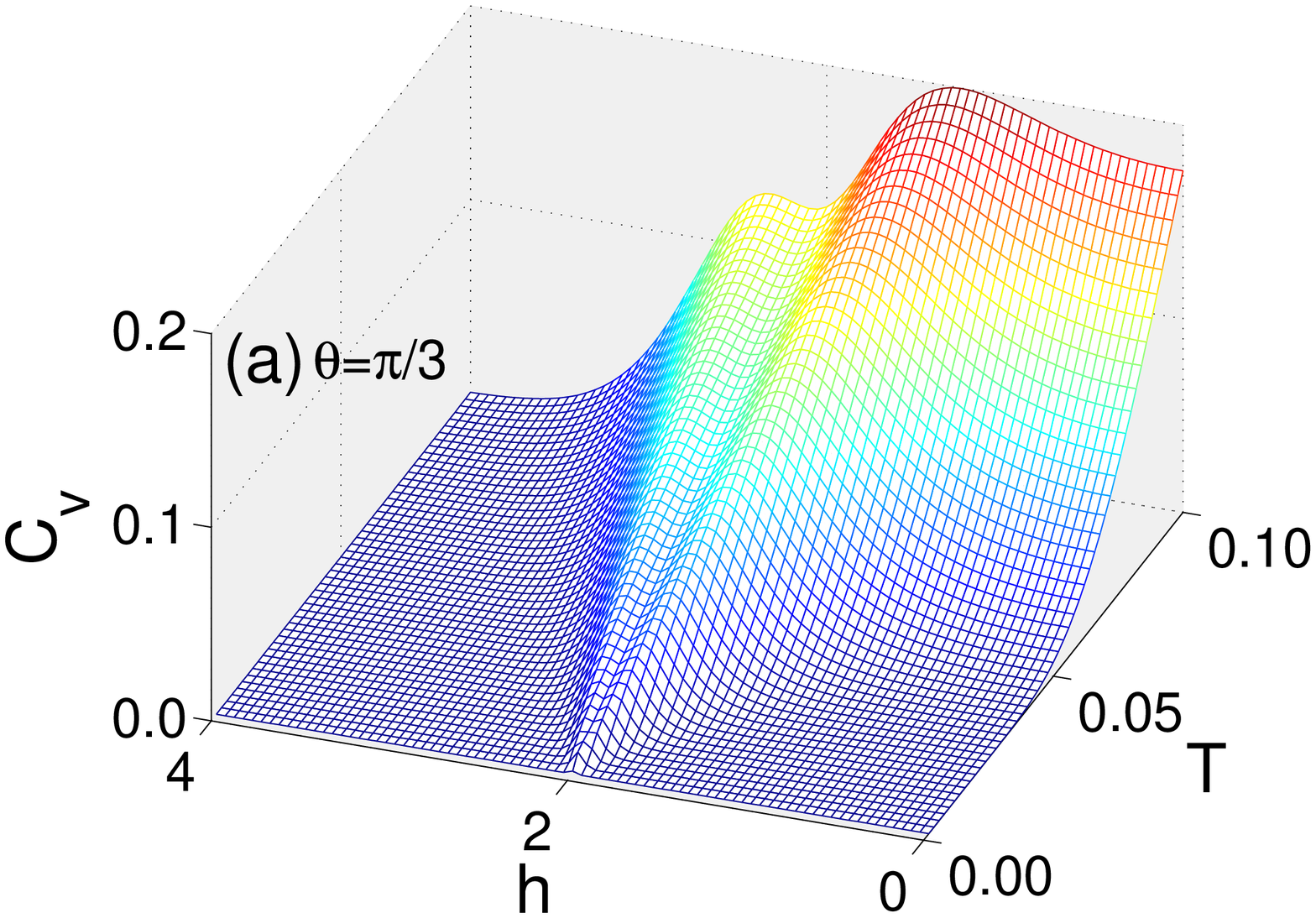}
\includegraphics[width=8cm]{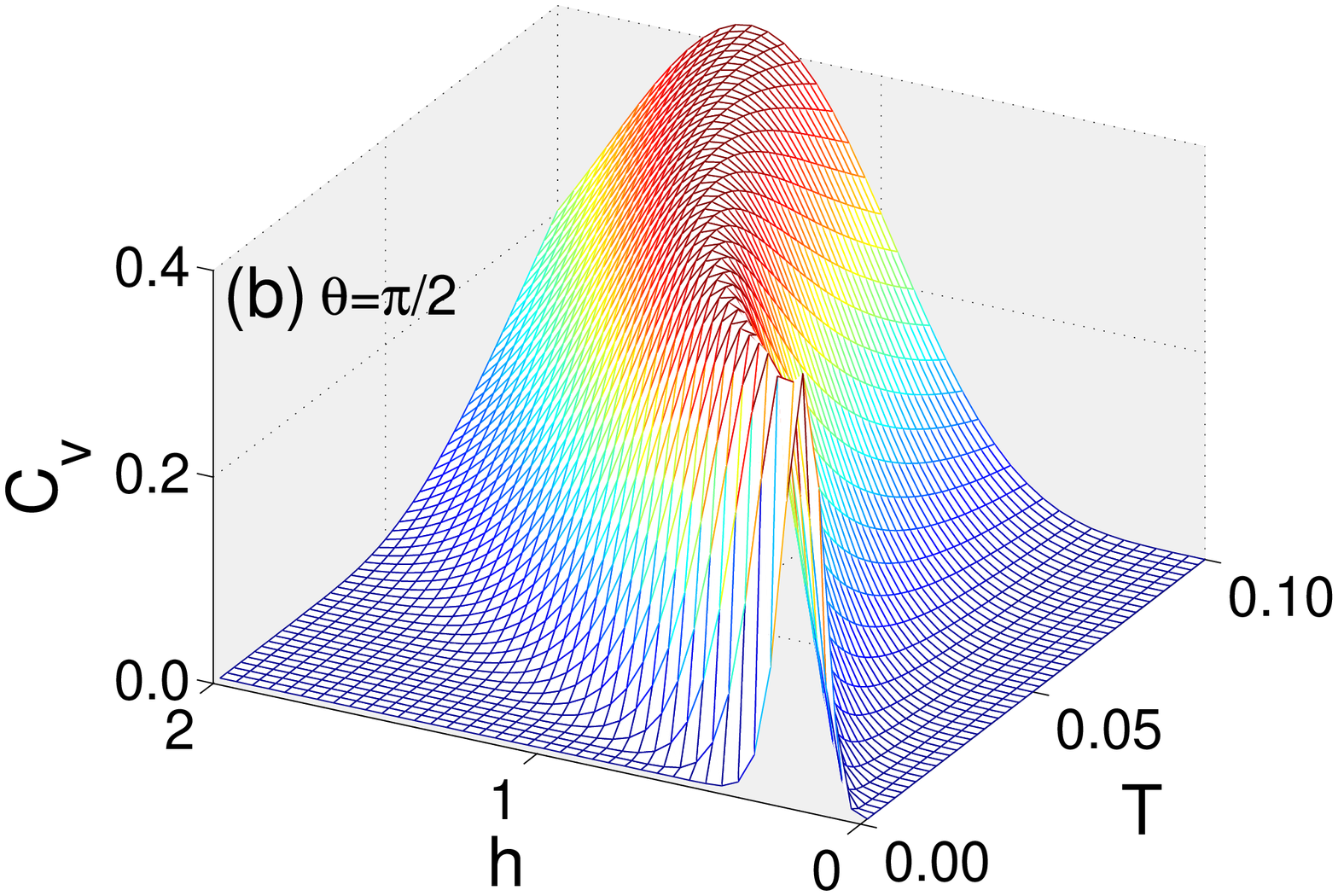}
\caption{(Color online)
The 3D plot of the specific heat $C_V$ normalized per unit cell for the:
(a) EOM at $\theta=\pi/3$, and
(b) QCM at $\theta=\pi/2$.
Note that the specific heat reaches its local minima at QCPs only
for extremely low temperatures.
Parameters: $J_{\rm o}$=1, $J_{\rm e}$=4. }
\label{fig:cv}
\end{figure}

\subsection{Specific heat for the 1D compass models}
\label{sec:cv}

Next we analyze the low temperature behavior of the heat capacity,
\begin{eqnarray}
\label{cv}
C_V&=&T\left(\frac{\partial{\cal S}}{\partial T}\right)_h
=\sum_k \sum_{j=1}^{2} \frac{\varepsilon_{k,j}^2}
{4T^2 \cosh^2 (\varepsilon_{k,j}/2T)}.
\end{eqnarray}
We recall that the entropy exhibits fast changes when the field $h$ is
close to its critical value, $h\approx h_c$ (but $h\neq h_c$), see Fig.
\ref{fig:ent}.
Here we concentrate on the qualitative differences between the EOM and
the QCM. The specific heat for both models is presented in 3D plots,
for increasing temperature and transverse field, see Fig. \ref{fig:cv}.
We have found that the low temperature behavior exhibits striking
differences between these models discussed below.

Consider first the EOM of Sec. \ref{sec:eg} [with angle $\theta=\pi/3$
in Eq. (\ref{gcom})]. The specific heat contains here a broad peak
around $h_c=2$ which corresponds to the QCP, and grows with increasing
temperature. This demonstrates that more entropy is released here, as
the spectrum of excited states is dense near $h\simeq h_c$. Furthermore,
$C_V$ develops a local minimum which splits the peak at $h\simeq h_c$
into two separate maxima for extremely low temperatures, see Fig.
\ref{fig:cv}(a). The maxima seen at $h<h_c$ and $h>h_c$ are of different
height which reflects the different spectra and increase of entropy with
increasing temperature in the vicinity of the QCP at $h_c=2$.
The shallow trough in heat capacity can be linked with orbital
susceptibility discussed in Sec. \ref{sec:chi} by the Maxwell relation
\cite{Jaf12}.

In contrast, increasing temperature at the QCP of the QCM ($h=h_c=0$)
does not result in any increase of the specific heat and one finds
$C_V=0$ in a broad range of temperature, see Fig. \ref{fig:cv}(b). This
somewhat surprising behavior is a consequence of the gap between the
excited states and the ground state. Here the ground state has high
macroscopic degeneracy, being $d=2^{N/2-1}$ --- this degenerate state
is a robust feature of the QCM, responsible for its rather unusual
properties, see also Sec. \ref{sec:chi}. Finite transverse field $h$,
however, splits the ground state multiplet, and the entropy at low
temperature decreases, see Fig. \ref{fig:ent}(b). Increasing temperature
for a constant but finite field $h$ results then in a fast increase of
entropy which is responsible for a large maximum in $C_V$ for the QCM,
as observed in Fig. \ref{fig:cv}(b).

\subsection{Orbital polarization and susceptibility}
\label{sec:chi}

In this Section we analyze the orbital properties at finite polarizing
field $h$ of both the EOM and QCM at finite temperature near the QPT.
From the free energy ${\cal F}$ we determined the orbital polarization
${\cal P}$ along the transverse field,
\begin{eqnarray}
{\cal P}=-\left(\frac{\partial{\cal F}}{\partial h}\right)_T
=\sum_k \sum_{j=1}^{2} \frac{\partial \varepsilon_{k,j}}{\partial h}
\tanh\left(\frac{\varepsilon_{k,j}}{2T} \right),
\end{eqnarray}
which vanishes as $h\to 0$. Thus, there is no polarization at any
finite temperature in one dimension and no nontrivial critical point,
in accordance with the Mermin-Wagner theorem. Also, there are no
peculiarities of the order parameter $\langle\sigma^z\rangle$ at any
finite temperature and finite transverse field $h$.

\begin{figure}[t!]
\includegraphics[width=8.2cm]{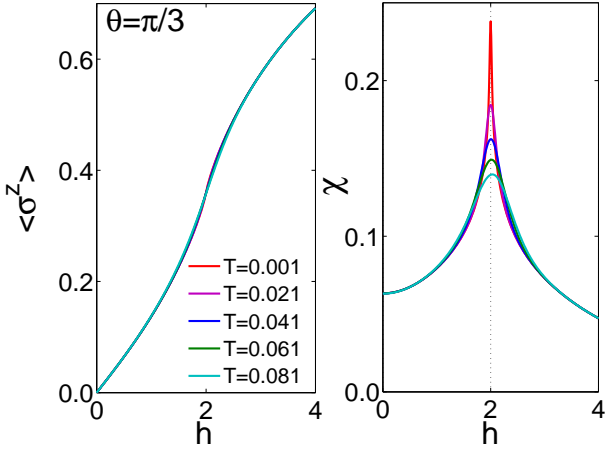}
\caption{(Color online) Orbital response in a transverse field $h$ for
the EOM at different temperatures:
(a) the orbital polarization $\langle\sigma^z\rangle$ per site, and
(b) the orbital susceptibility $\chi$ per site (\ref{chi}).
The QPT is found at $h_c=2$.
Different curves from top to bottom correspond to increasing
temperature and are normalized per one site.
Parameters: $J_{\rm o}$=1, $J_{\rm e}$=4, $\theta=\pi/3$. }
\label{fig:eom}
\end{figure}

\begin{figure}[b!]
\includegraphics[width=8.2cm]{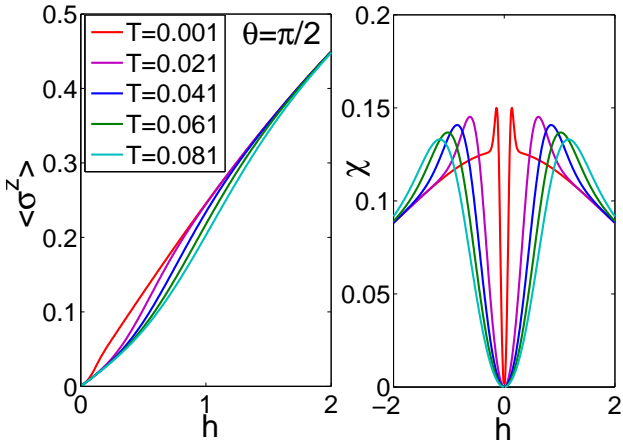}
\caption{(Color online) Orbital response in a transverse field $h$ for
the QCM at different temperature:
(a) polarization $\langle\sigma^z\rangle$ per site, and
(b) the orbital susceptibility $\chi$ per site (\ref{chi}).
Different curves from the center to the left or right correspond to
increasing temperature and are normalized per one site.
Parameters: $J_{\rm o}$=1, $J_{\rm e}$=4, $\theta=\pi/2$. }
\label{fig:qcm}
\end{figure}

The orbital susceptibility is the derivative of the polarization
${\cal P}$ (\ref{m}) over the field $h$, and we define it here per one
site,
\begin{equation}
\label{chi}
\chi=\frac{1}{N}
 \left(\frac{\partial        {\cal P}        }{\partial h}\right)_T
=\left(\frac{\partial\langle\sigma_i^z\rangle}{\partial h}\right)_T.
\end{equation}
After using the Jordan-Wigner fermions one finds it in the fermionic
representation,
\begin{eqnarray}
\label{chif}
\chi&=&\frac{1}{2N}\sum_k\sum_{j=1}^{2}
\left\{ \frac{\partial^2\varepsilon_{k,j}}{\partial h^2}
\tanh\left( \frac{\varepsilon_{k,j}}{2T}\right) \right. \nonumber \\
&+& \left.\left(\frac{\partial\varepsilon_{k,j}}{\partial h}\right)^2
\left[2T\cosh^2\left(\frac{\varepsilon_{k,j}}{2T}\right)\right]^{-1}\right\}.
\end{eqnarray}
We emphasize that the orbital properties (similar to magnetic
properties in spin models) are intimately related to
the field dependence of the entropy via the Maxwell identity,
\begin{equation}
\label{max}
 \left(\frac{\partial{\cal S}}{\partial h}\right)_T
=\left(\frac{\partial{\cal P}}{\partial T}\right)_h,
\end{equation}
which allows to rewrite the cooling rate as
\begin{equation}
\label{gammanew}
\Gamma_h=\frac{1}{C_V}\left(\frac{\partial{\cal P}}{\partial T}\right)_h.
\end{equation}
Therefore, we discuss below the orbital properties from the
perspective of the peculiarities of the entropy at finite field
and finite temperature, presented in Sec. \ref{sec:s}.

The polarization $\langle\sigma^z\rangle$ of the EOM increases with
field $h$ and this increase is almost independent of temperature except
in the vicinity of the phase transition, see Fig. \ref{fig:eom}(a).
At the critical field $h_c=2$ the derivative of the polarization
diverges at $T=0$, and in the low temperature regime one finds a sharp
maximum in the susceptibility $\chi$ at $h=h_c$, see Fig.
\ref{fig:eom}(b). This behavior represents a generic QPT with $h$ as
control parameter. We note that the associated peak in the entropy
leading to the phase transition is described here by the vanishing of
the gap $\Delta$ Eq. (\ref{gapdelta}) that occurs in the fermionic
spectrum at $h=h_c$.

\begin{figure}[t!]
\includegraphics[width=8cm]{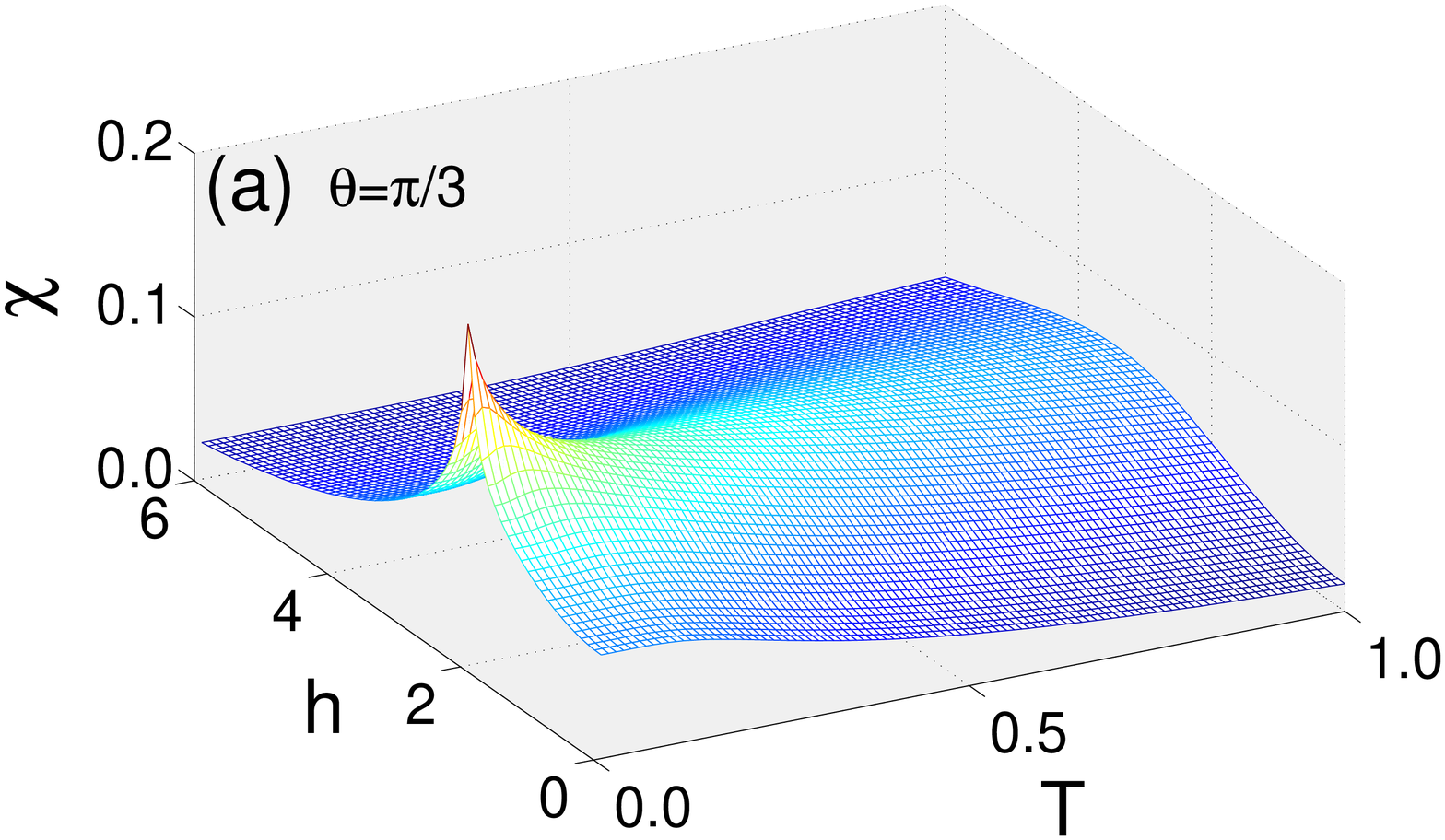}
\includegraphics[width=8cm]{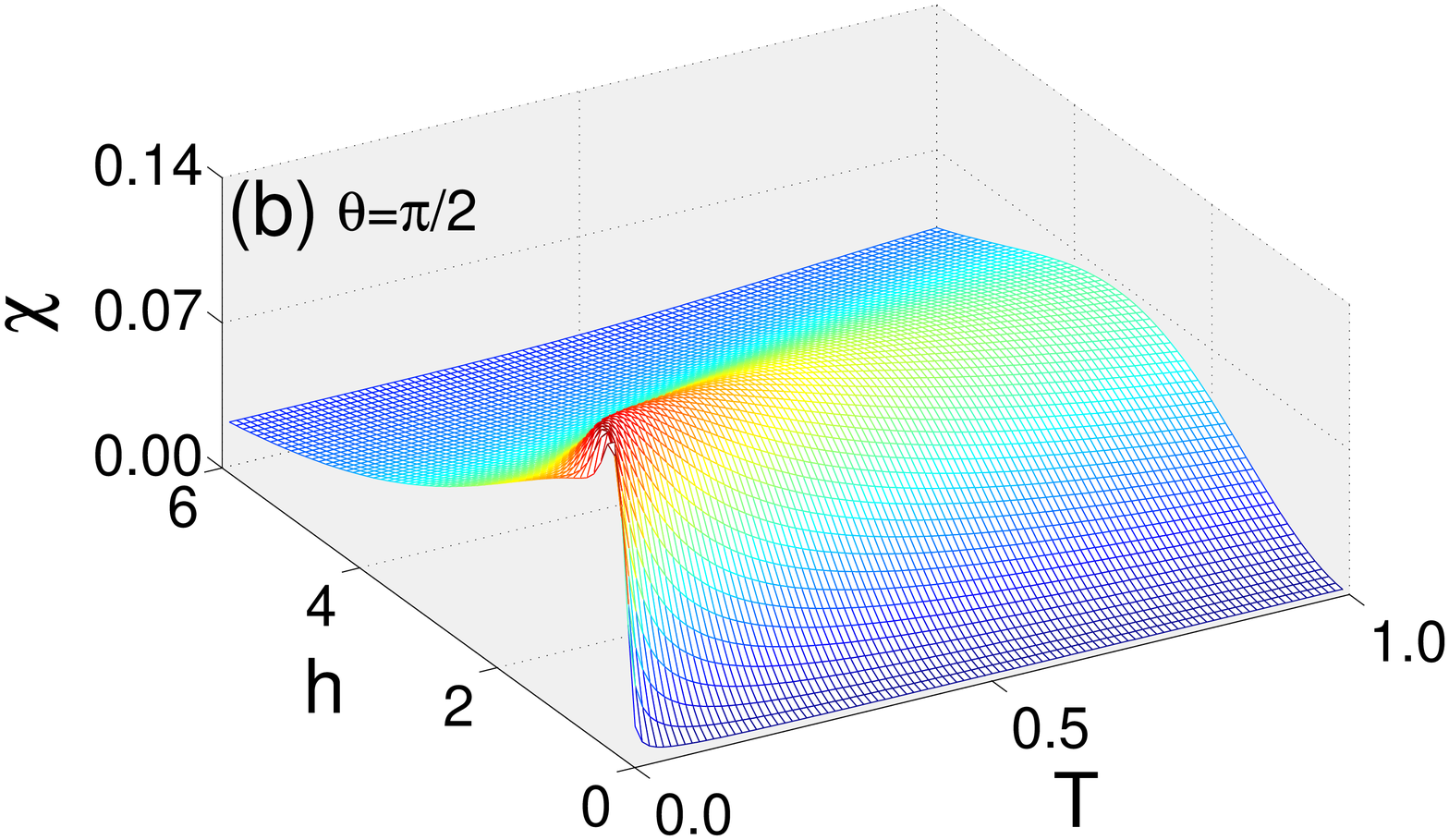}
\caption{(Color online) The 3D plot of the orbital susceptibility
$\chi$ (\ref{chi}) versus temperature and field for the:
(a) EOM at $\theta=\pi/3$, and
(b) QCM at $\theta=\pi/2$.
Parameters: $J_{\rm o}$=1, $J_{\rm e}$=4.
The EOM (a) shows a QPT at finite field $h_c=2$.
The QCM (b) is characterized by the macroscopic degeneracy of the low
energy sector and vanishing $\chi$ at zero field, while finite $h$
lifts the degeneracy and leads to a peak in $\chi$. }
\label{fig:chi}
\end{figure}

The 1D QCM shows a remarkably different orbital response. Here the
polarization increase with $h$ depends strongly on temperature, see Fig.
\ref{fig:qcm}(a). A clearer presentation of this peculiar behavior is
possible in terms of the susceptibility $\chi$, Fig. \ref{fig:qcm}(b).
Here $\chi(T,h)$ vanishes at $h=0$ and acquires a peak at a finite field
$h_m(T)$ which increases with temperature. This is another manifestation
of the macroscopic entropy at zero temperature, shown in Fig.
\ref{fig:ent}(b), that stems from the highly degenerate ground state.
In the fermionic language the vanishing of $\chi$ at $h=0$ is connected
with the high degeneracy of the subspace described by two dispersionless
half-filled  fermionic bands, $\pm\epsilon_{k,1}=0$. When the degeneracy
is lifted by a finite transverse field, the entropy changes dramatically
and causes a rapid increase of the susceptibility shown in Fig.
\ref{fig:qcm}(b). Below we shall discuss a different picture for the
origin of this degeneracy in the QCM.

Finally, we compare the orbital susceptibility $\chi$ (\ref{chif})
obtained for both 1D compass models (the EOM and the QCM) in a broad
range of temperature in Fig. \ref{fig:chi}. In the gapped phase of the
EOM at $h=0$, the low temperature orbital susceptibility is finite
and decreases with increasing $T$ for the unpolarized system,
see Fig. \ref{fig:chi}(a). In contrast, one finds a vanishing orbital
susceptibility at the critical point of the QCM $h=0$ in a broad range
of temperature. A distinct maximum develops close to $h=0$ at low
temperature --- this maximum moves to higher field and looses intensity
when temperature increases further, see Fig. \ref{fig:chi}(b).
All these distinct features emphasize once again radical difference
between the nature of the QPTs found in both compass models.

\section{Discussion and summary}
\label{sec:summa}

In this paper we explored the ground state and the thermodynamic
properties of the 1D generalized compass model with exchange
interactions given by Eq. (\ref{gJ}), and tuned by an angle $\theta$.
They vary from Ising interactions at $\theta=0$ to maximally frustrated
ones with two different pseudospin components coupled on even/odd bonds
at $\theta=\pi/2$ in the quantum compass model. In between
(at $\theta=60^\circ$) one finds the $e_g$ orbital model. In this way,
we investigated the consequences of increasing frustration of spin
interactions in one dimension. The model was solved exactly and we
presented its exact characteristics in the thermodynamic limit: the
entropy, the specific heat, the orbital susceptibility, and the
adiabatic demagnetization curves. By investigating the dependence of
all these quantities on the angle $\theta$, we have shown that the
ground state is ordered along the easy axis as long as
$\theta\neq\pi/2$, whereas it becomes disordered and highly degenerate
at $\theta=\pi/2$, i.e., when the
interacting pseudospin components along even/odd bonds are orthogonal.

Pseudospin excitations are separated by a gap from the ground state
everywhere except for the quantum compass model, where the gap closes
and one finds a highly disordered spin-liquid ground state.
This demonstrates an important difference between the $e_g$ orbital
model with favored type of short-range order and the quantum compass
model in one dimension. While the above order for the $e_g$ orbitals
is analogous to the 2D case \cite{Cin10}, the 1D compass model fails
to develop the nematic order known from its 2D analog.

\begin{figure}[t!]
\includegraphics[width=8cm]{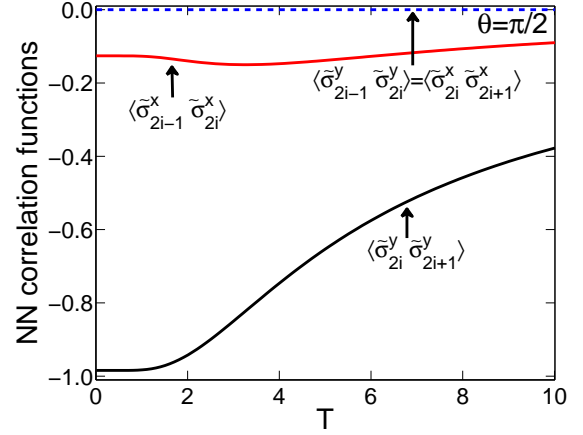}
\caption{(Color online) Nearest neighbor pseudospin correlations in
the anisotropic quantum compass model [the model Eq. (\ref{gcom})
with angle $\theta=\pi/2$ and stronger coupling of $\{\sigma_i^y\}$
components on the even bonds] for increasing temperature $T$.
Only the intersite correlation corresponding to the interacting
pseudospin components are finite.
Parameters: $J_{\rm o}=1$, $J_{\rm e}=4$. }
\label{fig:all}
\end{figure}

The generic temperature dependence of pseudospin correlations on the
bonds in the 1D quantum compass model is summarized in Fig.
\ref{fig:all}. Only these pseudospin correlations take finite values
which are coupled by finite interaction parameters, similar as in
the isotropic case \cite{Brz07}. As expected, at $T=0$ the value of
pseudospin correlation $|\langle\sigma_{2i}^y\sigma_{2i+1}^y\rangle|$ is
larger than $|\langle\sigma_{2i-1}^x\sigma_{2i}^x\rangle|$ as the first
one corresponds to a stronger interaction. On the other hand, the
complementary orbital correlations for pairs that are not coupled by
any interaction, i.e., $\langle\sigma_{2i}^x\sigma_{2i+1}^x\rangle$ and
$\langle\sigma_{2i-1}^y\sigma_{2i}^y\rangle$, vanish and this follows
from the $Z_2$ symmetry \cite{Che08}, as discussed also in the
Appendix. Note that a finite value of
$\langle\sigma_{2i-1}^x\sigma_{2i}^x\rangle$ in Fig. \ref{fig:all} is a
manifestation of the quantum nature of the compass model, as in the
classical case one finds instead only finite pseudospin correlations on
stronger bonds, i.e., $\langle\sigma_{2i}^y\sigma_{2i+1}^y\rangle=-1$
and $\langle\sigma_{2i-1}^x\sigma_{2i}^x\rangle= 0$.

Furthermore, we have shown that the external transverse field has also
quite different consequences, depending on the underlying interactions.
In the $e_g$ orbital model intersite pseudospin correlations are robust
and follow the strongest interactions. Therefore, a significant value
of the transverse field is required here to modify the short-range
correlations, dictated by the $\sigma_i^x\sigma_{i+1}^x$ interactions,
and to induce the polarized state. A qualitatively different situation
is encountered in the quantum compass model. Here the highly disordered
ground state is fragile and already an infinitesimal transverse field
destabilizes it and induces a quantum phase transition which
we recognized as being of second order by investigating the adiabatic
demagnetization at finite temperature. The observed behavior corresponds
to entropy maximization at the quantum critical point in the
low-temperature limit. The high degeneracy revealed by finite entropy
at low temperature suggests that the 90$^\circ$ compass model may have
potential application in quantum computation \cite{Rio11}. In addition,
the cooling rate in the quantum compass model could be testified
in the state-of-the-art experiments at optical lattice \cite{McK11}.

We would like to emphasize that some quantum integrable 1D models were
developed in the past to provide valuable insights into the nature of
quantum correlations in the ground state, as well as into the structure
of excited states. Such models:
(i) help to understand the nature of many-body states in such models,
and
(ii) provide a possibility to test approximate theoretical methods used
for more realistic physical models of frustrated spin interactions,
in two and three dimensions.
The present study should serve the same purpose.

Summarizing, we have demonstrated that robust pseudospin correlations
arise on the bonds in the $e_g$ orbital model --- these correlations
get destroyed only at the quantum phase transition which occurs at
rather strong transverse field.
On the contrary, the disordered spin-liquid state in the isotropic
quantum compass model is fragile and gets destroyed by infinitesimal
field. A qualitative difference is found for anisotropic interactions
--- the spin-liquid state is more robust here and survives up to
temperature $T^*$ which appears to be a new energy scale and increases
with increasing anisotropy of interactions. This feature follows from
the weak logarithmic decrease of spin entropy with increasing
temperature, and persisting high degeneracy of the ground state
in this temperature range.

\acknowledgments

We thank Wojciech Brzezicki for insightful discussions.
W.-L. Y. acknowledges support by the National Natural Science
Foundation of China (NSFC) under Grant No. 11004144.
A. M. O. acknowledges support by the Polish National Science
Center (NCN) under Project No. 2012/04/A/ST3/00331.

\appendix*
\section{Consequences of the $Z_2$ symmetry}
\label{sec:app}

In this Appendix we shall show that the macroscopic degeneracy of the
1D QCM which was manifested in two fermionic bands at zero energy,
$\pm\varepsilon_k=0$, is due to local $Z_2$ symmetries of the model in
the absence of the transverse field term. For this discussion we write
the QCM Eq. (\ref{gcom90}) in an equivalent form with simplified notation,
\begin{equation}
\label{qcmxz}
{\cal H}_{\rm QCM}=
-\sum_i(J_x \sigma^x_{2i-1}\sigma^x_{2i}+J_z\sigma^z_{2i}\sigma^z_{2i+1}),
\end{equation}
and introduce operators which act on bonds \cite{note1D}:
\begin{eqnarray}
\label{opx}
X_i&=& \sigma^x_{2i} \sigma^x_{2i+1},\\
\label{opz}
Z_i&=& \sigma^z_{2i-1} \sigma^z_{2i},
\end{eqnarray}
where each set of operators, $i=1,2,.....,N/2$, commutes with the Hamiltonian.
Thus we can use the tuple
\begin{equation}
\vec{Z}\equiv (Z_1,...,Z_i,....,Z_{N/2})
\end{equation}
to classify the eigenstates of ${\cal H}_{\rm QCM}$, for instance
\begin{eqnarray}
\label{ev}
{\cal H}_{\rm QCM} |\Phi_0\rangle &=& \varepsilon_0  |\Phi_0\rangle,\\
\vec{Z} |\Phi_0\rangle&=&(\lambda_1,...,\lambda_i, ....,\lambda_N)
|\Phi_0\rangle,
\end{eqnarray}
where the eigenvalues $\lambda_i=\pm 1$ follow from $Z_i^2=1$.
It is important to recognize that the operators $Z_i$ and $X_j$
anticommute for special cases:
\begin{eqnarray}
\label{XZ1}
\{Z_i,X_i\}&=&0,\\
\label{XZ2}
\{Z_i,X_{i-1}\}&=&\{Z_{i+1},X_i\}=0,
\end{eqnarray}
while they commute otherwise. We note here, that the key difference
to the 2D QCM \cite{Dou05,Dor05,Tro10} lies in the different form of these
anticommutation relations. Using the commutation relation
$[{\cal H}_{\rm QCM},X_i]=0$, one finds that
\begin{equation}
\label{XZ}
{\cal H}_{\rm QCM} X_i |\Phi_0\rangle = \varepsilon_0 X_i |\Phi_0\rangle,
\end{equation}
that is, also  $|\Phi_1\rangle = X_i|\Phi_0\rangle$ is an eigenvector
to the same eigenvalue $\epsilon_0$, and  by analyzing the corresponding
eigenvalue tupel $(\lambda_1,...,\lambda_i, ....,\lambda_{N/2})$ one can
convince oneself that this state is in fact distinct from
$|\Phi_0\rangle$. One can now proceed by applying the same arguments
to $|\Phi_2\rangle = X_{i-1} |\Phi_0\rangle$ and so on, until one
exhausts all the $2^{N/2}$ states of the degenerate multiplet.


\begin{thebibliography}{99}
\bibitem{notesz} We take here $\sigma_i^y$ instead of $\sigma_i^z$ spin
                 component in the EOM considered usually \cite{vdB99}.
\bibitem{note1D} Such bond operators were introduced as pseudospins to
                 classify the invariant subspaces in the 1D isotropic
                 compass model \cite{Brz07}.
\end{thebibliography}
\end{document}